\documentclass[%
 reprint,
 amsmath,amssymb,
 aps,
]{revtex4-2}

\usepackage{graphicx}
\usepackage{dcolumn}
\usepackage{bm}
\usepackage{xcolor}

\begin{document}

\preprint{APS/123-QED}

\title{Theoretical analysis of single-ion anisotropy in $d^3$ Mott  insulators}

\author{Xiaoyu Liu$^1$}
\author{Derek Churchill$^1$}
\author{Hae-Young Kee$^{1,2}$}
\email[]{hykee@physics.utoronto.ca}
\affiliation{Department of Physics, University of Toronto, Ontario, Canada M5S 1A7}
\affiliation{Canadian Institute for Advanced Research, CIFAR Program in Quantum Materials, Toronto, Ontario, Canada M5G 1M1 }




\date{\today}

\begin{abstract}
An effective spin model for Mott insulators is determined by the symmetries involved among magnetic sites, electron fillings, and their interactions. 
Such a spin Hamiltonian offers insight to mechanisms of magnetic orders and magnetic anisotropy beyond the Heisenberg model.
For a spin moment S bigger than 1/2, single-ion anisotropy is in principle allowed. 
However, for $d^3$ Mott insulators with large cubic crystal field splitting, the single-ion anisotropy is absent within the LS coupling, despite S = 3/2 local moment.
On the other hand, preferred magnetic moment directions in $d^3$ materials have been reported, which calls for a further theoretical investigation.
Here we derive the single-ion anisotropy interaction using the strong-coupling perturbation theory. 
The cubic crystal field splitting including $e_g$ orbitals, trigonal distortions,  Hund's coupling, and spin-orbit coupling beyond the LS scheme are taken into account.
 {For compressed distortion, the spin-orbit coupling at magnetic sites can favor either the easy-axis or the easy-plane while that of anions leads to easy-axis anisotropy.}
We apply the theory on $\rm{CrX}_3$ with X = Cl and I, and show the dependence of the single-ion anisotropy on  {the strength} of the spin-orbit couplings of both magnetic and anion sites. 
Significance of the single-ion anisotropy in ideal two-dimensional magnets is also discussed.
\end{abstract}

\keywords{Suggested keywords}
\maketitle

\section{Introduction}

Two-dimensional (2D) magnets have been of great interest in both fundamental and applied research communities due to their intrinsic long-range order (LRO)
and potential application in spintronics, data storage, and sensing~\cite{soumyanarayanan_emergent_2016,miao_2d_2018}. 
In particular, recent progresses on 2D materials such as monolayer $\rm{CrI}_3$~\cite{huang_bevin_2017} and bilayer $\rm{Cr}_2\rm{Ge}_2\rm{Te}_6$~\cite{gong_cr2ge2te6_2017} have generated intense theoretical and experimental activities
to understand and control physical properties via pressure, strain, doping, and/or stacking into heterostructures~\cite{huang_gating_CrI3_2018,sivadas_stacking-dependent_2018,webster_strain_CrX3_2018,wu_strain-tunable_CrI3_2019,cai_atomically_crcl3_2019,li_pressure_CrI3_2019}.
These 2D materials exhibit paramagnetic (PM) to ferromagnetic (FM) transition at a critical temperature $T_c$.
This immediately implies that their effective spin model is beyond SU(2) symmetric Heisenberg interaction, because 
there is no LRO in 2D Heisenberg magnets at any finite temperature due to thermal fluctuations, i.e, celebrated Mermin-Wagner theorem~\cite{mermin_wagner_1966}. 
Thus the magnetic anisotropy is crucial for 2D magnets to hold the LRO at finite temperatures. 
Previous studies showed that such anisotropy includes the single-ion anisotropy (SIA) for spin ${\bf S}$ bigger than 1/2, 
XXZ model~\cite{lado_XXZ_2017,kim_XXZ_2019}, and/or bond-dependent interactions such as Kitaev and $\Gamma$ interactions~\cite{xu_SIAK_2018,lee_KGammal_2020}, as they are allowed by the symmetry of crystal.

While the symmetry is a strong constraint to the effective spin model, it is not sufficient to determine the pinning of magnetic moment direction and the size of spin gap essential for a finite temperature LRO.  
To access the information beyond the symmetry-allowed terms, the spin Hamiltonian in relation to virtual hoppings between different magnetic sites is necessary. 
Such a model can be derived using the standard strong coupling expansion theory starting from the multi-orbital Kanamori-Hubbard interaction~\cite{kanamori_1963} and  treating inter- and intraorbital hoppings as  perturbations. 
It is well established that the magnetic anisotropy including popular bond-dependent Kitaev and $\Gamma$  interactions originates from the interplay between 
spin-orbit coupling (SOC), crystal field splitting as well as Hund's coupling~\cite{jackeli_mott_2009,rau_spinmodel_2014,kim_rucl3_2015,stavropoulos_S=1_2019, stavropoulos_2021}.

 {For $d^3$ Mott insulators such as Cr$^{3+}$, there are three electrons in six $t_{2g}$ orbitals in the limit when the cubic crystal field splitting is infinite (i.e., ignoring the $e_g$ orbitals). This maps to the half-filled $t_{2g}$ orbitals, where the total spin ${\bf S} = \sum_i {\bf s}_i = \frac{3}{2}$ and total angular momentum ${\bf L} = \sum_i {\bf l}_i = 0$ based on the first and second Hund’s rule respectively. 
In this case, the SIA is absent because ${\bf L}=0$ within the LS coupling scheme ($\lambda {\bf L}\cdot {\bf S}$). This means that the spin anisotropy should come from a finite trigonal crystal splitting and/or beyond the LS coupling, i.e., sum of each atomic SOC, $\xi \sum_i {\bf l}_i\cdot {\bf s}_i$. }
In real solid-state materials,  there is an additional crystal field splitting from trigonal distortion, as 2D materials are grown on substrates, which is crucial for a finite SIA in addition to SOC. While the above arguments are expected, the analytical expression of SIA for $d^3$ systems has not been fully explored.

In this paper, we study how the SIA depends on the SOC, crystal field splitting, Hund's coupling, and trigonal distortion in $d^3$ S = 3/2 systems. 
We present analytical expressions for SIA in various limits. We find the SIA depends on the relative strength of the cubic crystal field splitting and the Hund's coupling. The easy-axis versus easy-plane direction is determined 
by the trigonal distortion when the $e_g$ contribution is included, while they work against each other in the large Hund's coupling limit. 
 {For a compressed distortion, the SOC at magnetic sites can either favor the easy-plane or -axis depending on the $p$-$d$ hybridization, while that of anions leads to easy-axis single-ion anisotropy.
}
We hope our result will offer a useful guideline to estimate the SIA and enhance $T_c$ in $d^3$ systems.

The paper is organized as follows. In Sec. II, we discuss the onsite Hamiltonian and its spectrum under SOC and trigonal distortions. 
In Sec. III, we discuss the spin model for  $d^3$ S = 3/2, and the SIA from the strong-coupling perturbation method. In Sec. IV, we discuss the SIA originated from the $p$-orbital SOC. In Sec.~V, we apply our theory on $\rm{CrX}_3$, with X=Cl, I,  and show how the total SIA from both the magnetic and anion sites depends on the relative strength of SOC between them.
A short summary and discussion are presented in the last section. 

\section{\label{sec:onsite}The onsite Hamiltonian}
\begin{figure}
    \centering
    \includegraphics[scale=0.8]{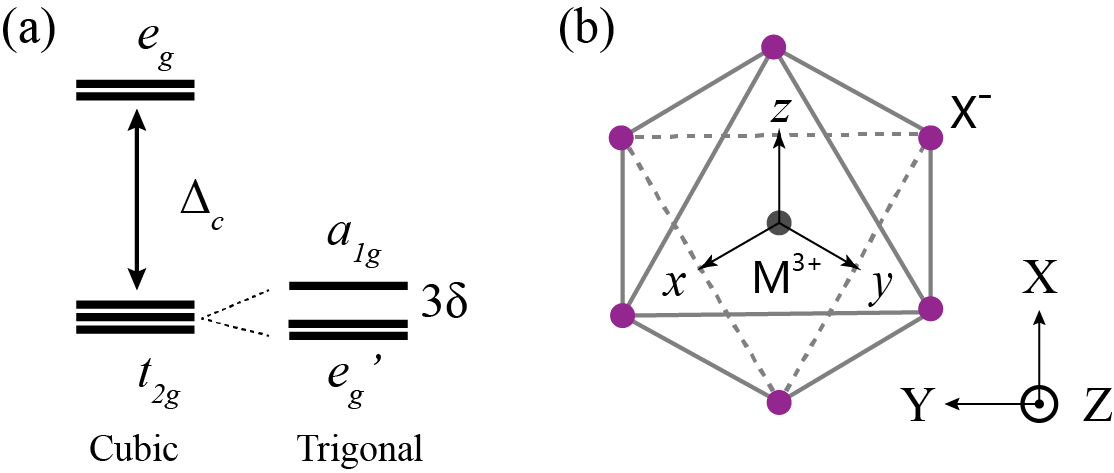}
    \caption{ {(a) Crystal field splitting under cubic and trigonal fields. The splitting between $a_{1g}$ and $e'_g$ is $3\delta$ where $\delta$ denotes the trigonal field effect defined in Eq. (\ref{eq:trig}). (b) An octahedral cage in MX$_3$. The local axis ($x,y,z$) is defined along the octahedral axis. The global axis ($X,Y,Z$) is defined with $Z$ along the (111) direction in the local coordinate system, perpendicular to the edge-sharing octahedral honeycomb. }}
    \label{fig:illustration}
\end{figure}

MX$_3$ where M a transition metal and X a halide is composed of edge-sharing MX$_6$ octahedra, forming a 2D honeycomb structure. The octahedral coordination of the MX$_6$ cages leads to a cubic crystal field splitting (CFS) $H_{\rm{cubic}}=\sum_{\alpha\in e_g}\Delta_c\  c_{\alpha}^\dagger c_{\alpha}$ on the M site, as shown in Fig. \ref{fig:illustration}(a). 
Beside the cubic CFS, in van der Waals materials, the octahedral cages are usually trigonally distorted, leading to a further trigonal field splitting $\delta$ shown in Fig. \ref{fig:illustration}(a) with
\begin{equation}\label{eq:trig}
H_{\rm{trig}}=\left(
\begin{array}{ccc}
    0 & \delta & \delta  \\
    \delta & 0 & \delta\\
    \delta & \delta & 0
\end{array}
\right) 
.
\end{equation}
The equation is written in basis ($d_{xy}$,$d_{yz}$,$d_{zx}$) , where the $x$,$y$ and $z$ are the local axes of the octahedron, as shown in Fig. \ref{fig:illustration}(b). It is equivalent to $H_{\rm{trig}}=\delta (2-3L_Z^2)$~\cite{huimei_prl_2020}
with $L_Z$ being the angular momentum along the $Z$ direction which is perpendicular to the 2D honeycomb lattice, as shown in Fig. \ref{fig:illustration}(b).  Compression of the octahedral cage prefers $L_Z=0$ which is generally associated with positive $\delta$. 

Since we are interested in the effective spin model of multi-orbital Mott insulators, we begin with the Kanamori-Hubbard model~\cite{kanamori_1963}.
\begin{equation}
\begin{aligned}\label{eq:kanamori}
H_{\rm{Coulomb}} &= U\sum_\alpha n_{\alpha\uparrow} n_{\alpha\downarrow} +
\frac{U'}{2} \sum_{\alpha\neq\beta,\sigma,\sigma'} n_{\alpha \sigma} n_{\beta \sigma'}\\
&-\frac{J_H}{2}\sum_{\alpha\neq\beta,\sigma\sigma'}c_{\alpha\sigma}^\dagger c_{\beta\sigma'}^\dagger c_{\beta \sigma} c_{\alpha \sigma'}\\
&+ J_H\sum_{\alpha\neq\beta} c_{\alpha\uparrow}^\dagger c_{\alpha\downarrow}^\dagger c_{\beta \downarrow} c_{\beta \uparrow},
\end{aligned}
\end{equation}
where the $U$ and $U'$ are the intra and interorbital Coulomb interactions, $J_H$ is the Hund's coupling. $c^\dagger_{\alpha \sigma}$ and $c_{\alpha \sigma}$ are creation and annihilation operators of $\alpha$ orbital with spin $\sigma$. $n_{\alpha \sigma}$ is the density operator.

 {Here we use the simplified multi-orbital model ignoring 3- and 4-orbital interaction terms, which become important when $e_g$ orbitals are not well separated from $t_{2g}$~\cite{sugano_multiplets_2014,coury_hubbard_2016,wang_edrixs_2019}. Since the cubic crystal field splitting is rather large, we expect that the simplified Kanamori Hamiltonian Eq. (\ref{eq:kanamori}) is a good approximation. We indeed find including 3- and 4-orbital interaction terms, which is denoted by “full” interaction model in Appendix \ref{sec:full_coulomb}, gives small corrections to the SIA. }

Without SOC the spins do not have a preferred direction within spin space.
To explain the (intrinsic) magnetic anisotropy in MX$_3$ systems, we include the SOC to entangle the spin and orbitals defined on a lattice. 
The atomic SOC is given by the summation of the SOC on each electron $i$,
\begin{equation}\label{eq:lisi}
H_{\rm{SOC}} = \xi_M\sum_i \bf{l}_i \cdot \bf{s}_i.
\end{equation}
Here the $\bf{l}_i$ and $\bf{s}_i$ are the angular momentum and spin momentum of each electron respectively. 
The SOC effect can be approximated by $H_{\rm{SOC}}^{\rm{LS}}=\lambda(L,S) \bf{L}\cdot \bf{S}$ as discussed earlier, 
and we will consider the both cases and show how the results of SIA differ between the two approaches.

The total onsite Hamiltonian is the summation of the above terms 
\begin{equation}\label{eq:htot}
H_{\rm{tot}}=H_{\rm{Coulomb}}+H_{\rm{cubic}}+H_{\rm{SOC}}+H_{\rm{trig}}.
\end{equation}

When $\Delta_c > U$, the electrons on M$^{3+}$ ions tend to stay on $t_{2g}$ orbitals. 
When the Hund\rq{}s coupling is finite, the lowest energy state is described by three electrons in $t_{2g}$  aligned to form S = 3/2.
The excited states depend on the strength of the cubic CFS $\Delta_c$. When $\Delta_c$ is infinite,  where $e_g$ orbitals are not taken into account,
we will be limited to all excited states within $t_{2g}^3$ configurations. 
Throughout this paper we will use $t_{2g}^3$ for such a case where $e_g$ states are not considered, while we will use $d^3$ for three electrons in any $d$-orbitals in excited states. 
Without SOC and trigonal field splittings, the spectrum of $t_{2g}^3$ is listed in the first part of Table \ref{table:spectrum}. 
The lowest fourfold degenerate states have total spin $S=\frac{3}{2}$ and total angular momentum $L=0$. The two sets of excited states are 10-fold and sixfold with $L=2$, $S=\frac{1}{2}$ and $L=1$, $S=\frac{1}{2}$ respectively. 

\begin{figure}
    \centering
    \includegraphics[width=1\columnwidth]{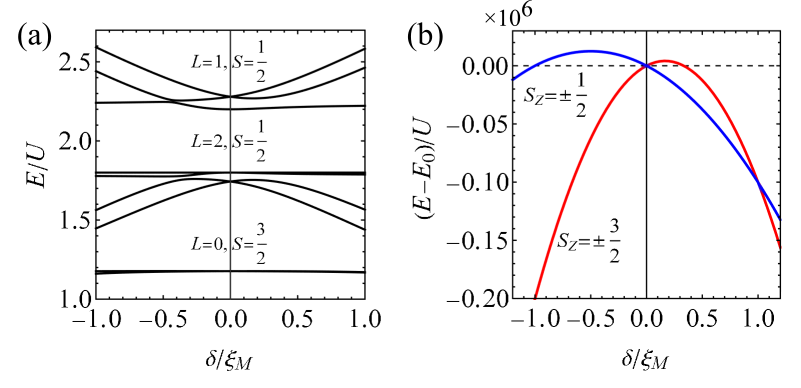}
    \caption{(a) Spectrum of $t_{2g}$ with $J=0.2U$, $\xi_M=0.15U$. $\xi_M$ is enlarged to enlarge the splittings. (b) Lowest states with $\xi_M=0.01U$. See also ~\cite{suzuki_spin_2019}}
    \label{fig:t2gspectrum}
\end{figure}

In the presence of SOC and trigonal field, couplings between the lowest states and the excited states are enabled. 
Carrying out the numerical calculations for small SOC and trigonal field, the spectrum under this circumstance as a function of $\delta/\xi_M$ is shown in Fig. \ref{fig:t2gspectrum}(a). 
When we zoom in to the lowest states,  we find the lowest fourfold degenerate $S=\frac{3}{2}$ states split into two doublets with $S_Z=\pm\frac{1}{2}$ (blue) and $S_Z=\pm\frac{3}{2}$ (red), as shown in Fig. \ref{fig:t2gspectrum}(b). 
For small positive $\delta$ (corresponding to compression along $Z$ direction), $S_Z=\pm\frac{1}{2}$ doublets have lower energy, indicating a preference of spin moment lying in-plane. 
Around $\delta\approx\xi_M$ for a positive $\delta$, the two doublets cross again and $S_Z=\pm\frac{3}{2}$ are preferred at large positive $\delta$,  consistent with the earlier numerical result found in ~\cite{suzuki_spin_2019}.

The above finding is under the assumption of $t_{2g}^3$ configuration. 
In real materials, excited states can have electrons in any $d$ orbitals including the $e_g$ orbitals, i.e., $d^3$ configuration. 
The exact spectrum of $d^3$ configuration cannot be obtained analytically, as the Hund's coupling and the cubic CFS do not commute with each other. 
Thus we present $d^3$ spectrum under two extreme conditions, as shown in the second and third parts of Table \ref{table:spectrum}. 
In the limit $\Delta_c =0$, there are 40-fold, 70-fold and 10-fold degenerate states with energy $3U-9J_H$, $3U-6J_H$ and $3U-2J_H$ respectively. 
On the other hand, when $J_H = 0$, there are 20-fold, 60-fold, 3sixfold and fourfold degenerate states with energy $3U$, $3U+\Delta_c$, $3U+2\Delta_c$ and $3U+3\Delta_c$ respectively, depending on the number of electrons in $e_g$ orbitals. 

\begin{figure}
    \centering
    \includegraphics[width=1\columnwidth]{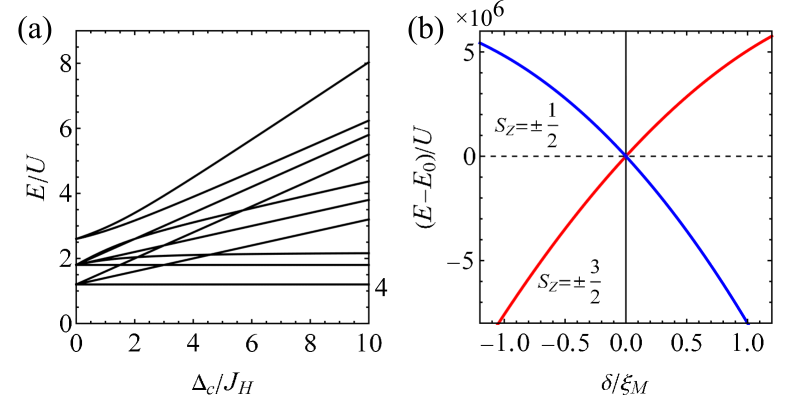}
    \caption{(a) Spectrum of $d^3$ without SOC and trigonal distortions. The fourfold degeneracy of the lowest states is labeled.  (b) Splitting of $d^3$ lowest states. We set $J_H=0.2U$, $\Delta_c=0.3U$, $\xi_M=0.01U$.}
    \label{fig:d3spectrum}
\end{figure}

For finite $\Delta_c$ and $J_H$, $d^3$ spectrum as a function of $\Delta_c/J_H$  is obtained numerically as shown in Fig. \ref{fig:d3spectrum}(a). 
With any finite $\Delta_c$, 40-fold degenerate states split and the lowest states are given by fourfold S=3/2 states as expected. 

Similar to the above discussion, including SOC and trigonal distortions can also lead to couplings between S=3/2 states and higher states, leading to splittings of the S=3/2 quadruplets as shown in Fig. \ref{fig:d3spectrum}(b) for a given ratio of $\Delta_c/U = 0.3$ and $J_H/U = 0.2$.
The splitting between $S_Z = \pm \frac{3}{2}$ and $S_z = \pm \frac{1}{2}$ is larger than $t_{2g}^3$ case, while the tendency of having $S_Z =\pm \frac{1}{2}$ for positive $\delta$ is also found without crossing around $\xi_M \sim \delta$. 
The larger splitting in $d^3$ than $t_{2g}^3$ indicates that $e_g$ orbitals in excited states are important and 
their contribution dominates the SIA strength.
Furthermore,  for a positive $\delta$,  $S_z = \pm \frac{1}{2}$ is always lower in energy, implying the easy-plane SIA. 
Below we will perform the strong coupling perturbation theory to obtain the analytic expressions of the SIA in two cases, $t_{2g}^3$ and $d^3$. 

\begin{table}[h]
\caption{\label{table:spectrum}
Spectrum. Assume $U'=U-2J_H$ }
\begin{ruledtabular}
\begin{tabular}{cc}
Degeneracy & Energy \\\hline
\multicolumn{2}{c}{$t_{2g}^3$}\\
4  & $3U-9J_H$ \\
10 & $3U-6J_H$ \\
6  & $3U-4J_H$ \\\hline
\multicolumn{2}{c}{$d^3$,$\Delta_c=0$}\\
40 & $3U-9J_H$\\
70 & $3U-6J_H$\\
10 & $3U-2J_H$\\\hline
\multicolumn{2}{c}{$d^3$,$J_H=0$}\\
20 &  $3U$\\
60 &  $3U+\Delta_c$\\
36 &  $3U+2\Delta_c$\\
4 &  $3U+3\Delta_c$\\
\end{tabular}
\end{ruledtabular}
\end{table}

\section{Analytical derivation of single ion anisotropy}
Based on symmetry, the low-energy effective spin model for S=3/2 using the octahedra coordinate system $x-y-z$ is given by~\cite{stavropoulos_2021}

\begin{equation}\label{eq:spin_ham}
\begin{aligned}
H_{\rm{spin}} = & \sum_{\langle i j \rangle\in \alpha\beta(\gamma)}
J\textbf{S}_i\cdot\textbf{S}_j+
K S_i^\gamma S_j^\gamma+
\Gamma (S_i^\alpha S_j^\beta+S_i^\beta S_j^\alpha)\\
& +\Gamma' (S_i^\alpha S_j^\gamma + S_i^\beta S_j^\gamma + S_i^\gamma S_j^\alpha + S_i^\gamma S_j^\beta)\\
& +A ({\bf S}_i \cdot {\hat Z})^2.
\end{aligned}
\end{equation}

Among them,  it was shown that $\Gamma$ is absent up to the fourth order perturbation term, while $\Gamma^\prime$ is introduced by the trigonal distortion.
Here in this paper we focus on the SIA term (last term) of the above spin model. 
$S_i^Z$ is the spin component at site $i$ along $Z$, see Fig. \ref{fig:illustration}(b). The coefficient $A>0$ corresponds to easy-plane and $A<0$ easy-axis. 
In Sec. II,  we have shown how the energy spectra split due to the trigonal and SOC numerically.
In this section, we derive analytically the expression of $A$ by using strong-coupling perturbation  theory~\cite{stavropoulos_2021}. 
Both $\textbf{l}_i \cdot \textbf{s}_i$ and $\textbf{L}\cdot\textbf{S}$ SOC schemes are considered.

To derive the spin model, we start from eigenstates of 
$H_0=H_{Coulomb}+H_{cubic}$ and treat $H_{SOC}$ and $H_{trig}$ as perturbation $V \equiv H_{SOC}+H_{trig}$. 
The total  Hamiltonian can be written in the subspace of lowest states of S=3/2 and the subspace of excited states as follows. 
\begin{equation}\label{eq:perturbham}
    H=
    \left(
    \begin{array}{cc}
        H_0  &   \\
         & H_1  \\
    \end{array}
    \right)+
    \left(
    \begin{array}{cc}
        V_{00} & V_{01}  \\
        V_{10} & V_{11}  \\
    \end{array}
    \right).
\end{equation}
where the subscripts $0$ and $1$ refers to the lowest and excited states, respectively. 
Using downfold technique, 
an effective Hamiltonian is then given by
\begin{equation}\label{eq:downfold}
H_{\rm{eff}}= H_0+V_{01}\frac{ 1}{E_0-H_1-V_{11}}V_{10}.\\
\end{equation}
When ${\rm min}(E_0-H_1)$ is greater than $V_{11}$, we can further expand the fraction as
\begin{equation}\label{eq:taylor_series}
\frac{ 1}{E_0-H_1-V_{11}}=\frac{1}{E_0-H_1}+\frac{1}{E_0-H_1} V_{11} \frac{1}{E_0-H_1}+....\\
\end{equation}

Below we show the results of SIA for different limits. 

\subsection{$t_{2g}^3$ when $\Delta_c\rightarrow\infty$}
Without $e_g$ orbitals, the eigenspace of $H_0$ and its energy spectra are listed in Table \ref{table:spectrum} where  4, 10 and 6-dimensional subspaces are classified by 
$L$ and $S$ within the LS coupling scheme. 
The total Hamiltonian in these 4, 10, and 6 degenerate basis is written as  
\begin{equation}\label{eq:t2g3_20_LS}
\begin{aligned}
H = \left(
    \begin{array}{ccc}
        0  &  &  \\
         & 3J_H & \\
         &  & 5J_H
    \end{array}
    \right)+
    \left(
    \begin{array}{ccc}
        0_{4\times 4}  & 0_{4\times10} & 0_{4\times6} \\
        0_{10\times 4} &H'_{11}(\lambda) & H'_{12}(\delta)\\
        0_{6\times4} & H'_{21}(\delta) & H'_{22}(\lambda)
    \end{array}
    \right)
\end{aligned}.
\end{equation}
where from the perturbation part, we find that the fourfold lowest states are decoupled from the excited states. Thus the SIA under LS coupling is zero as expected due to the quenched angular momentum in $t_{2g}^3$. 

Beyond the LS coupling, consider the SOC given by  $\xi_M \sum_i {\bf l}_i\cdot {\bf s}_i$, the perturbation part is
\begin{equation}\label{eq:t2g3_20}
\begin{aligned}
H' = 
    \left(
    \begin{array}{ccc}
        0_{4\times 4}  & 0_{4\times10} & H'_{02}(\xi_M) \\
        0_{10\times 4} & 0_{10\times 10} & H'_{12}(\xi_M,\delta)\\
        H'_{20}(\xi_M) & H'_{21}(\xi_M,\delta) & 0_{6\times6}
    \end{array}
    \right)
\end{aligned},
\end{equation}
The dependence of trigonal distortion is the same as Eq. (\ref{eq:t2g3_20_LS}) as we write the Hamiltonian in the same basis. 
However, contrary to Eq. (\ref{eq:t2g3_20_LS}), $\xi_M$ dependent $H'_{02}$ and $H'_{12}$ are non-zero. 
$H'_{02}(\xi_M)$ connects the lowest fourfold subspace with the excited states. This matrix structure indicates that the SIA under ${\bf l}_i\cdot {\bf s}_i$ coupling is finite.

We find that up to fourth order perturbation theory, the SIA is given by
\begin{equation}\label{eq:Ac_t2g3}
    A_M(t_{2g}^3)=\frac{\delta \xi_M^2(\xi_M-\delta)}{25J_H^3}.
\end{equation}
The subscript $M$ indicates that the SIA is induced by the SOC of the M-site. The sign of $A_M(t_{2g}^3)$ is determined by the sign of $\delta$ and the relative strength of $\delta$ and $\xi_M$. 
This behavior is consistent with the numerical result shown in Fig. \ref{fig:t2gspectrum}(b) where the sign change of  {$A_M$} occurs around $\delta \sim \xi_M$. 
The difference between these two SOC schemes is due to the fact that the LS coupling is an approximation of ${\bf l}_i \cdot {\bf s}_i$ by treating the SOC 
between the LS subspaces as a perturbation and keeping only the diagonal elements with the lowest order~\cite{fazekas_1999}. 

\subsection{$d^3$ including $e_g$ orbitals}
As shown in Sec. III, when the $e_g$ orbitals are included, the spectrum of $d^3$ is significantly different from $t_{2g}^3$. 
The spectrum of $d^3$ is rather complicated with the presence of both Hund's coupling $J_H$ and cubic CFS $\Delta_c$. 
We obtained the expression of SIA in the 120-dimensional $d^3$ space by the similar method described in the last subsection. 
We find that within the LS coupling scheme,  the SIA is given by  
\begin{equation}\label{eq:Ac_d3_LS}
    A_M^{LS}=\frac{6\delta\lambda^2}{\Delta_c^2}
\end{equation}
On the other hand,  using the $\xi_M \sum_i {\bf l}_i \cdot {\bf s}_i$ coupling,  SIA is found as
\begin{equation}\label{eq:Ac_d3}
    A_M=\frac{2}{3}\delta\xi_M^2 \left(\frac{1}{\Delta_c^2}-\frac{1}{(\Delta_c+3J_H)^2}-\frac{6}{(10\Delta_c+21J_H)^2}\right).
\end{equation}

There are several implications.
Firstly, it is well-known~\cite{fazekas_1999} that the relation between coefficient of $\lambda$ of ${\bf L}\cdot {\bf S}$ coupling and coefficient 
$\xi_M$ of ${\bf l}_i\cdot {\bf s}_i$ is $\lambda=\pm\xi_M/(2S)$ with positive corresponding to less than half-filled and negative for more than half-filled. 
According to the above relation, $\lambda=\xi_M/3$ for $d^3$ configuration. Substituting this relation into Eq. (\ref{eq:Ac_d3_LS}), we find it is exactly the first term of Eq. (\ref{eq:Ac_d3}), while the second and third terms are beyond the LS scheme.

Secondly, we notice that the dominant contribution to SIA in Eq. (\ref{eq:Ac_d3}) is the first term
which originates from the excitations to $e_g$ orbitals.  The details can be found in Appendix \ref{Sec:d3_40}.
The Hund's coupling gives negative corrections, reducing the SIA strength. 
When $J_H$ becomes tiny, $A_M$ becomes negative.
However, the local moment of S=3/2 requires a finite $J_H$ and we expect the positive $A_M$ favoring $S_Z = \pm \frac{1}{2}$ when $\delta$ is positive. 

Lastly, comparing with the SIA of the $t_{2g}^3$ case where a finite SIA occurs at the fourth order (see Eq. (\ref{eq:Ac_t2g3})), the SIA for $d^3$ is a third order term.  
Thus the contribution from $e_g$ orbitals dominate the SIA strength. 
This is consistent with the numerical results of the energy splittings between $S_Z=\pm\frac{3}{2}$ and $S_Z=\pm\frac{1}{2}$  shown in Fig. \ref{fig:t2gspectrum}(b) and Fig.  \ref{fig:d3spectrum}(b).  
Also the linear dependence of $\delta$ is consistent with Fig. \ref{fig:d3spectrum}(b). For compression (positive $\delta$), the SIA from the combination of SOC and trigonal distortion at magnetic site  always prefer easy-plane anisotropy. 

The SIA for $t_{2g}^3$ and $d^3$ under different SOC coupling schemes is summarized in Table \ref{table:Ac_analytical}. These are shown to the lowest order of SIA for each case. 

\begin{table}[h]
\caption{Analytical expression for SIA} \label{table:Ac_analytical}
\begin{ruledtabular}
\begin{tabular}{ccc}
& $\lambda L\cdot S$ & $\xi_M {\bf l}_i\cdot {\bf s}_i$  \\\hline
$t_{2g}^3$& 0 & $\frac{\delta\xi_M^2(\xi_M-\delta)}{25J_H^3}$ \\
$d^3$& $\frac{6\delta\lambda^2}{\Delta_c^2}$ & $\frac{2}{3}\delta\xi_M^2(\frac{1}{\Delta_c^2}-\frac{1}{(\Delta_c+3J_H)^2}-\frac{6}{(10\Delta_c+21J_H)^2})$
\end{tabular}
\end{ruledtabular}
\end{table}

The summary shown in Table \ref{table:Ac_analytical} indicates that the SOC at magnetic sites with  positive trigonal distortion leads to an easy-plane (positive $A$) SIA. On the other hand, several MX$_3$ reports easy-axis (negative $A$) SIA, which should come from beyond the on-site contribution to SIA. Below we investigate the contributions from the anions via hopping processes.

\subsection{Contributions from anion SOC}
Aside from the above onsite contribution to the SIA, the SOC on anions also contributes to SIA through distortion induced hoppings ~\cite{stavropoulos_2021}. 
A rigorous derivation of $A$ should include full processes including hopping between $M$ and $X$ sites involving charge configurations such as $d^4p^5$.
For simplicity, here we use an effective hopping model derived from integrating out the hopping to anions.
Up to linear order of distortion induced hoppings, we found the SIA is given by


 {
\begin{equation}\label{eq:Ac_intersite}
\begin{aligned}
A_X=& -\left(\frac{4}{5J_H}+\frac{16}{5(10\Delta_c+21J_H)}\right) \frac{t_A}{t_\pi} t_{\rm eff}^2\\
&+\frac{6J_H}{\Delta_c(\Delta_c+3J_H)} \frac{t_\sigma (t_\pi t_B+ t_\sigma t_C)}{t_\pi^3} t_{\rm eff}^2\\
\end{aligned}
\end{equation}}
where the subscript $X$ indicates SIA induced by SOC on an $X$ site. 
 {The effective hopping is given by $t_{\rm eff} = \frac{2t_\pi^2}{3} \left( \frac{1}{\Delta_{pd}-\frac{\xi_X}{2}} - \frac{1}{\Delta_{pd}+\xi_X}\right)$}.

The distortion-induced hoppings are parameterized as shown in the Appendix \ref{sec:wannier} and  {$t_A=-2\delta t_1+\delta t_2+\delta t_3+\delta t_4+\delta t_5$} and  {$t_B=\frac{1}{\sqrt{3}}(\delta\tau_1+2\delta\tau_2-\sqrt{3}\delta\tau_3$)} as well as $t_{C}=\delta t_6 +\delta t_7$. 

\section{Application to  ${\rm CrX}_3$ with ${\rm X= Cl}$ and I}
Here we apply our theory to 3$d^3$ CrX$_3$, since SOC and trigonal distortion are smaller than other energy scales.
To determine all necessary parameters such as $\Delta_c$,  $\delta$, and hopping parameters, 
we perform density functional theory (DFT) calculations. 
DFT calculations are performed with Vienna \textit{ab initio} Simulation Package (VASP)~\cite{vasp1993} without the Coulomb interaction and SOC. 
The projector augmented wave (PAW)~\cite{paw1994} potential and Perdew-Burke-Ernzerhof (PBE)~\cite{pbe1996} exchange-correlation functional are used.
 {
The experimental structures~\cite{morosin_xray_1964,braekken_kristallstruktur_1932,mcguire_CrI3_2015} are fully relaxed with SOC and various values of Hubbard $U$ ranging from 0 eV to 4 eV until the force on each atom is less than  0.01 eV/\AA. We find the structures for different $U$ values are very similar. In the following discussion, we use the relaxed structure with $U$=4 eV as an example.
For both the relaxation and static calculation, we use an energy cutoff of 350 eV and a $7\times7\times7$ k-point mesh.}
The tight-binding parameters are obtained from Wannier90 code~\cite{wannier90_2020}.
 {The Wannier parameters are listed in Appendix \ref{sec:wannier}}
The atomic SOC parameters within DFT without correlations are computed using the SOC matrix elements of a single atom in a 20~\AA$\times$20~\AA$\times$20~\AA box 
by OPENMX~\cite{ozaki_openmx_2003,ozaki_openmx_2004}. The atomic SOC for Cr, Cl, Br, and I are 31 meV, 82 meV, 326 meV and 646 meV respectively.

 {
Table \ref{table:dft} shows the effective $\delta$ and $\Delta_c$  from the Wannier model after downfolding into the $d$-orbitals. Since the cubic and trigonal crystal field splittings strongly depend on the $p$-$d$ hybridization and the underlying lattice structures, we first relax the bulk Cr$X_3$ structures. The values listed in Table \ref{table:dft} are then obtained within LDA using the relaxed structures without $U$ and $J_H$. The trend from $X$ = Cl to I is clear. While $\Delta_c$ and $\delta$  decrease, $\xi_X$ increases. It is important to note that $\delta$ changes sign for CrI$_3$ after downfolding, indicating the importance of $p$-$d$ hybridization (before the downfolding it is positive like Cl and Br; see the Appendix \ref{sec:wannier} for details). This means that both $A_M$ and $A_X$ are negative for the I case, leading to easy-axis anisotropy, while for Cl and Br, the opposite contributions to SIA from $A_M$ and $A_X$ occur.}

\begin{figure}
    \centering
    \includegraphics[width=0.95\columnwidth]{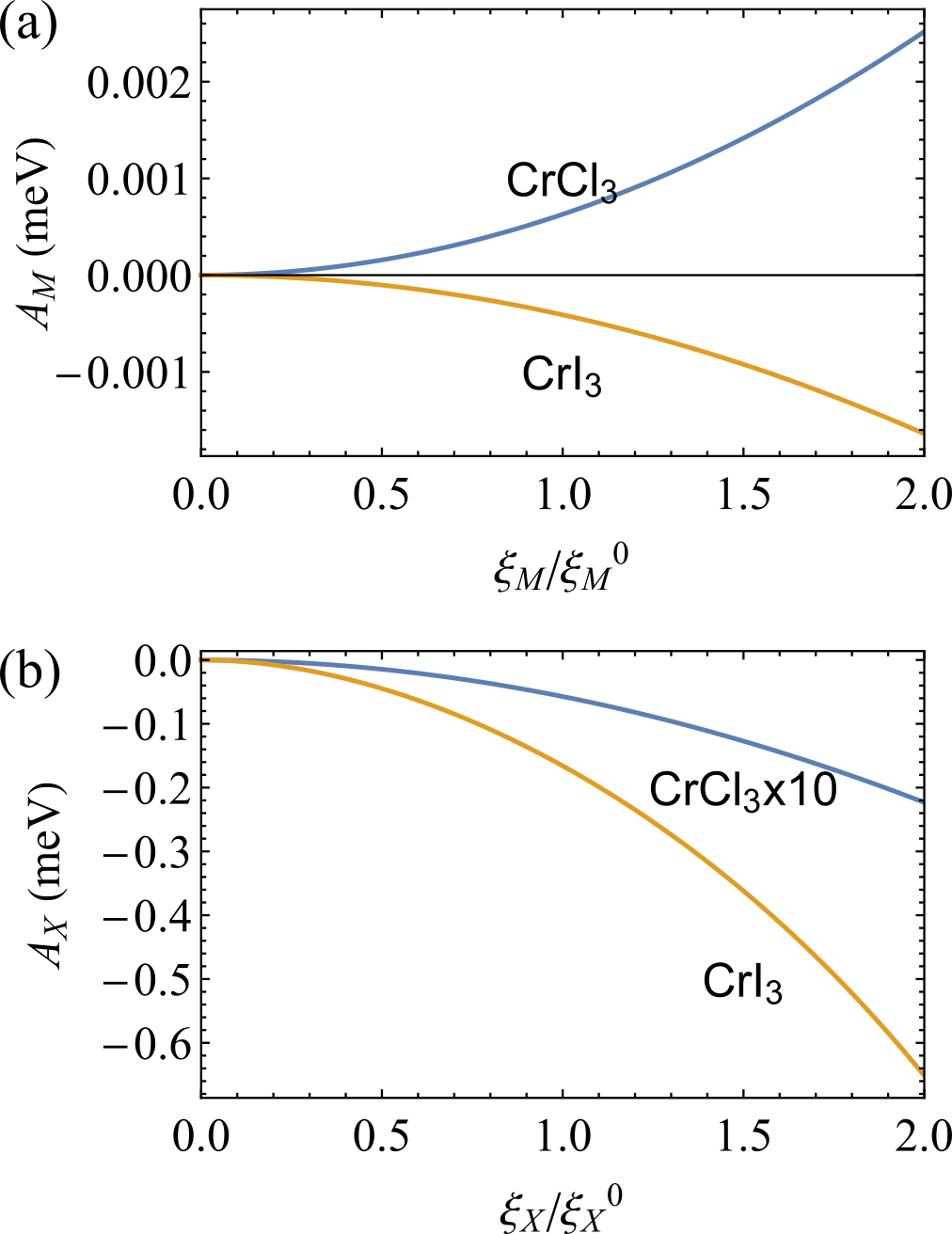}
    \caption{ {SIA arises (a) from M-site SOC $A_M$  given by Eq. (\ref{eq:Ac_d3}) and (b) from X-site SOC $A_X$ given by Eq. (\ref{eq:Ac_intersite}) with respect to the relative SOC strength. $J_H$ is chosen to be 1 eV. They are both  quadratic in SOC. The sign difference in $A_M$ for both materials is due to the sign difference in total effective trigonal field $\delta$ given in Table \ref{table:dft}. 
    }}
    \label{fig:Anumerical}
\end{figure}

 {Since the $M$ and $X$ site SOC may have opposite contributions and their strength can be enhanced by the electron-electron correlations ~\cite{kim_rucl3_2015,isobe_enhancement_2015,tamai_SRO_2019}, we leave $\xi_M$ and $\xi_X$ as two variables, and plot the SIA strength $A_M$ as a function of $\xi_M$ and similarly $A_X$ as a function of $\xi_X$ for CrCl$_3$ and CrI$_3$ for a fixed $J_H$ = 1 eV as shown in Fig. \ref{fig:Anumerical}. We find $A_X$ being negative for both, while $A_M$ is positive for CrCl$_3$ but negative for CrI$_3$. The sign change in $A_M$ in CrI$_3$ is due to the sign change of $\delta$ via $p$-$d$ hybridization as mentioned above. }

 {Experiments ~\cite{cable_CrCl3_1961,mcguire_CrCl3_2017} reported that CrCl$_3$ has moments lying in the plane, while CrI$_3$ has moments out of plane ~\cite{dillon_CrI3_1965,mcguire_CrI3_2015}. Given that the calculated $A_M$ is not large enough to compensate $A_X$ for CrCl$_3$, we speculate that the effective SOC at $M$ site could be further enhanced by electron-electron correlations ~\cite{kim_rucl3_2015,isobe_enhancement_2015,tamai_SRO_2019}, which remains for a future study. On the other hand, for CrI$_3$ due to the negative sign of $\delta$, the total SIA from both $A_M$ and $A_X$ is always negative leading to the easy-axis anisotropy. Quantifying the trigonal field strength is a challenging task, as it depends on the details of $p$- and $d$-orbital hybridization and corresponding charge densities. We note that the current work does not aim to offer precise values of SIA in CrX$_3$, but to provide the understanding of the SIA originated from the different combinations of SOC and trigonal field in $d^3$ systems.}


\begin{table}[h]
\caption{ \label{table:dft}
 {DFT parameters in meV. $\Delta_c$ and $\delta$ are obtained from the Wannier model with $d$ orbitals which takes into account $p$-orbital hybridization. $\Delta_{pd}$ is obtained from the $pd$ Wannier model. } 
}
\begin{ruledtabular}
\begin{tabular}{cccccccc}
& $\Delta_{pd}$ & $\Delta_c$ &    $\delta$ & $\xi_X$ \\\hline
$\mathrm{CrCl}_3$ & 2851 & 1481  & 2.45 & 82\\
$\mathrm{CrBr}_3$ & 2476 & 1329  & 0.80 & 326 \\
$\mathrm{CrI}_3$  & 2080 & 1169 & -0.96 & 646 \\
\end{tabular}
\end{ruledtabular}
\end{table}


\section{Discussion and Summary }

The existence of ferromagnetic LRO in two-dimensional (2D) systems with higher transition temperature $T_c$ has attracted intense studies.
To achieve a higher $T_c$ in ideal 2D materials, it is essential to have a certain magnetic anisotropy that opens up a spin gap which allows the system to
avoid quantum fluctuations and set up a LRO at finite temperature.  Thus understanding a microscopic origin of magnetic anisotropy in two-dimensional single-layer 
will guide ways to move towards a higher $T_c$.  While the full analysis of factors that determines $T_c$ is beyond the scope of the current study, as 
it requires a higher stiffness not only a finite spin gap,  our study will offer valuable inputs to the current efforts of enhancing $T_c$. 
 
In summary, we have studied a microscopic route to the SIA for S=3/2 in $d^3$ Mott insulator starting from the Kanamori-Hubbard interaction including Hund\rq{}s coupling, and
take into account the  CFS,  SOC,  and trigonal distortion.  We found that $e_g$ orbitals contribution is essential to understand the SIA strength and that the tendency towards easy-plane versus easy-axis is determined by two contributions denoted by $A_M$ and $A_X$.
 {For compressed trigonal distortion, the SOC at the magnetic sites can choose either easy-plane or easy-axis depending on the sign of $\delta$. When $\delta > 0$, it prefers the easy-plane, while $\delta <0$ easy-axis. The sign of $\delta$ is determined by the metal-ligand hybridization, and we found that for CrCl$_3$, it takes a positive value, while for CrI$_3$, it is negative leading to the easy-axis anisotropy. 
On the other hand, for $A_X$, it prefers the easy-axis for both Cr trihalides.}

Since we have used both SOC and trigonal distortion smaller than Hund\rq{}s coupling,  this theory is more applicable to $3d^3$ than $5d^3$ systems,
where $J_{\rm eff} = 3/2$ may be a better starting point than S=3/2 spin states. 
Recent works~\cite{badrtdinov_mop_3sio_11_2021,maharaj_La2LiOsO6_2018,Kermarrec_ba2yoso6_2015} of $5d^3$ have shown that in these systems there is a large spin gap. We propose that this may be relevant to the atomic SOC discussed in this paper. However cubic materials have very little distortions, implying possible bond-dependent interactions generated by SOC.  Extending the current theory to the stronger SOC may explain the anisotropy observed in these systems, which is a project for future studies. 

\begin{acknowledgments}
this paper was supported by the Natural Sciences and Engineering Research
Council of Canada and the Canada Research Chairs Program. This research was enabled in part by support provided by Sharcnet (www.sharcnet.ca) and Compute Canada
(www.computecanada.ca). Computations were performed on the GPC and Niagara super-
computers at the SciNet HPC Consortium. SciNet is funded by: the Canada Foundation
for Innovation under the auspices of Compute Canada; the Government of Ontario; Ontario
Research Fund - Research Excellence; and the University of Toronto.
\end{acknowledgments}

\appendix

\section{Coulomb interaction with 3- and 4-orbital effect}\label{sec:full_coulomb}

\begin{figure}
    \centering
    \includegraphics[width=0.9\columnwidth]{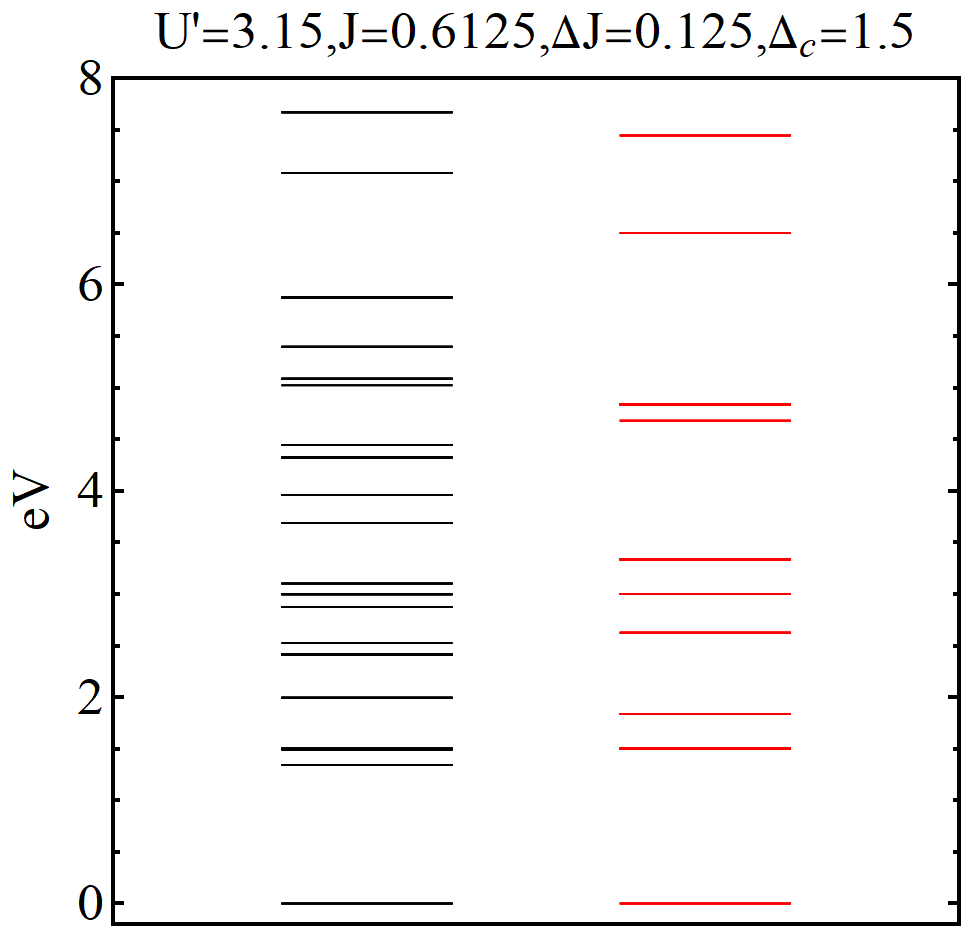}
    \caption{Spectrum of 3 electrons on 5-$d$ orbitals using both full interaction Hamiltonian including 3- and 4- orbital effects (Black, see Eq. (2) and (20) of ~\cite{coury_hubbard_2016}) and simplified multi-orbital Hamiltonian (Red, see Eq. (\ref{eq:kanamori}). }
    \label{fig:coury}
\end{figure}

\begin{figure}
    \centering
    \includegraphics[width=0.9\columnwidth]{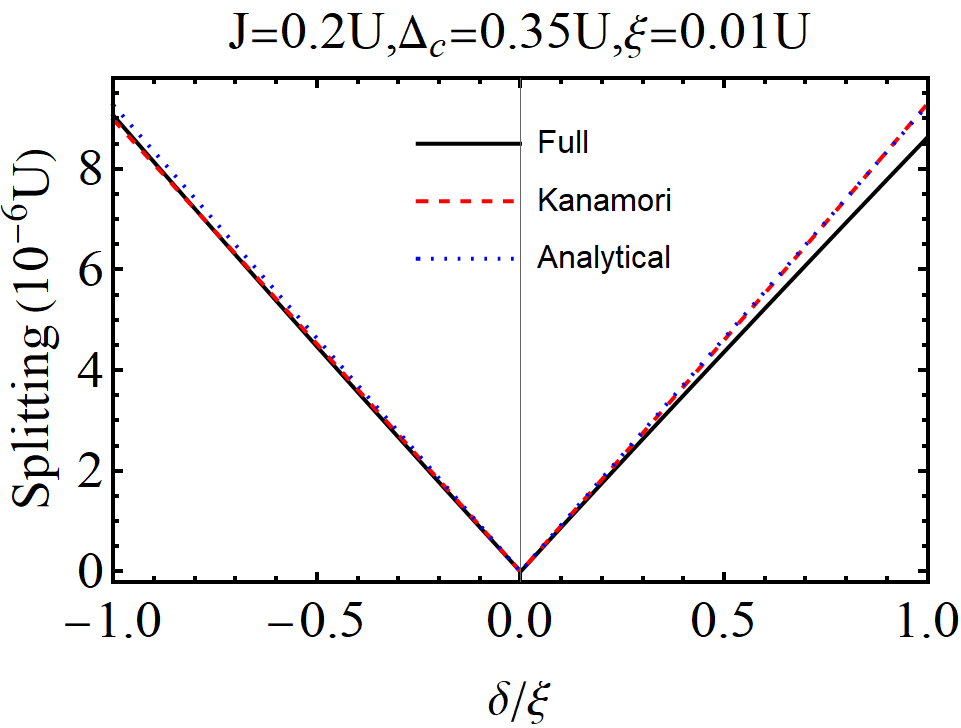}
    \caption{Energy splitting between the $S=\pm\frac{3}{2}$ doublets and $S=\pm\frac{1}{2}$ doublets under full multi-orbital  Hamiltonian and simplified Hamiltonian as well as the analytical expression from Eq. (\ref{eq:Ac_d3}). }
    \label{fig:comparison}
\end{figure}

 {
To include the missing 3- and 4-orbital interaction terms, we compared our Eq. (\ref{eq:kanamori}) with Eq. (20) of  ~\cite{coury_hubbard_2016}. We find by setting $\Delta J = 0$, the latter reduces to the former. 
The spectra of 3-electron many-body states for both cases are compared in Fig. \ref{fig:coury}. Here we set the crystal field splitting $\Delta_c = 1.5 {\rm eV }$ to be consistent with our DFT results. Though they are different in high energy range, they have similar lower excited states. The energy splittings of $S_Z=\pm\frac{3}{2}$ and $S_Z=\pm\frac{1}{2}$ with and without 3- and 4-orbital interactions are compared in Fig. \ref{fig:comparison}. The full form, the simplified one and the analytical result (Eq. (\ref{eq:Ac_d3})) are all very consistent with each other very well. This shows that the 3- and 4-orbital effects only have minor quantitative corrections to our results. }

\section{Downfolding}\label{Sec:downfold1}
The perturbation process in Sec. III can also be written in another way of infinite expansion.
If we write the Hamiltonian in more than two subspaces 
\begin{equation}\label{eq:perturbham1}
    H=
    \left(
    \begin{array}{ccc}
        H_0  & &  \\
         & H_1  &\\
         & & ...\\
    \end{array}
    \right)+
    \left(
    \begin{array}{ccc}
        V_{00} & V_{01} &... \\
        V_{10} & V_{11} & ...\\
        ... & ...& ...\\
    \end{array}
    \right),
\end{equation}
the effective Hamiltonian can be expanded as 
\begin{equation}\label{eq:downfold1}
\begin{aligned}
H_{eff}=& \sum_{i}V_{0i}\frac{ 1}{E_0-H_i-V_{ii}}V_{i0}\\
& +\sum_{i\neq j}V_{0i}\frac{1}{(E_0-H_i-V_{ii})}V_{ij}\frac{1}{(E_0-E_j-V_{jj})}V_{j0}\\
& +...\\
=&\sum_i V_{0i}\frac{1}{E_0-H_i} V_{i0}
\\
& +\sum_{i j}V_{0i}\frac{ 1 }{(E_0-H_i)}V_{ij} \frac{1}{(E_0-H_j)}V_{j0}+...\\
=& \sum_{i\neq j, j\neq k}\frac{H'_{0i}H'_{ij}H'_{jk}H'_{k0}}{(E_0-E_i)(E_0-E_j)(E_0-E_k)}.
\end{aligned}
\end{equation}
This expansion is equivalent to Eq. (\ref{eq:downfold}).

\section{$A$ analysis}\label{Sec:d3_40}
For $d^3$ configuration,  keeping only the excited states with energy $\Delta_c$ and $2\Delta_c$ (in other words we work in the 40-fold subspace where $\Delta_c\rightarrow0$). The Hamiltonian is
\begin{equation}\label{eq:d3_40}
        H=
    \left(
    \begin{array}{ccc}
        0 & &  \\
         & \Delta_c & \\
         & & 2\Delta_c
    \end{array}
    \right)+
    \left(
    \begin{array}{ccc}
        0_{4\times4} & H'_{01}(\xi_M) & 0_{4\times12}  \\
        H'_{10}(\xi_M) & H'_{11}(\xi_M,\delta) & H'_{12}(\xi_M)  \\
        0_{12\times4} & H'_{21}(\xi_M) & H'_{22}(\xi_M,\delta)  \\
    \end{array}
    \right).
\end{equation}
Up to third order perturbation  theory, we have 
\begin{equation}\label{eq:Ac_d3_40}
    A_c=\frac{2\delta\xi_M^2}{3\Delta_c^2}.
\end{equation}
This gives the first term in Eq. (\ref{eq:Ac_d3}) and is dominant as discussed in the main text. 

\section{ {Wannier models}}\label{sec:wannier}
 {We have two ways of building Wannier tight-binding models. One with only the $d$-orbitals and the other with both $d$ and $p$ orbitals. The former is effectively integrating out the $p$ orbitals in the latter due to the strong $p$-$d$ hybridization in this material. This strong hybridization can dramatically change the cubic CFS $\Delta_c$ as well as the trigonal CFS $\delta$ as can be read out from the following parameters. } 
\subsection{$d$-only wannier model}
 {
The onsite Hamiltonians are (written in sequence of $d_{x^2-y^2}$, $d_{3z^2-r^2}$,$d_{yz}$,$d_{xz}$,$d_{xy}$ and in unit meV): 
\begin{equation}
       H^{\textrm{CrCl}_3}= \left(
    \begin{array}{ccccc}
3790.08 & 0 & -4.01 & 3.43 & 0.58\\
0 & 3790.09 & 1.64 & 2.64 & -4.34\\
-4.01 & 1.64 & 2309.22 & 2.44 & 2.45\\
3.43 & 2.64 & 2.44 & 2309.22 & 2.45\\
0.58 & -4.34 & 2.45 & 2.45 & 2309.17\\
    \end{array}\right),
\end{equation}
\begin{equation}
       H^{\textrm{CrBr}_3}= \left(
    \begin{array}{ccccc}
3255.98 & 0 & -8.17 & 9.54 & -1.36\\
0 & 3255.98 & 6.28 & 3.92 & -10.27\\
-8.17 & 6.28 & 1926.85 & 0.79 & 0.81\\
9.54 & 3.92 & 0.79 & 1926.84 & 0.8\\
-1.36 & -10.27 & 0.81 & 0.8 & 1926.81
    \end{array}\right),
\end{equation}
\begin{equation}
       H^{\textrm{CrI}_3}= \left(
    \begin{array}{ccccc}
5141.62 & 0 & -12.43 & 10.31 & 2.12\\
0 & 5141.61 & 4.72 & 8.38 & -13.17\\
-12.43 & 4.72 & 3972.61 & -0.96 & -0.96\\
10.31 & 8.38 & -0.96 & 3972.61 & -0.95\\
2.12 & -13.17 & -0.96 & -0.95 & 3972.59
    \end{array}\right).
\end{equation}
The difference between diagonal terms of $e_g$ and $t_{2g}$ orbitals gives $\Delta_c$ listed in Table \ref{table:dft}. 
}

\subsection{$pd$ wannier model}
 {
The Hamiltonian in this $pd$ model can be written as 
\begin{equation}\label{eq:pdwannier}
       H = \left(
    \begin{array}{cc}
H_{\rm M} & T_{\rm MX}\\
T_{\rm MX}^T & H_{\rm X}
    \end{array}\right).
\end{equation}
Here the $H_{\rm M}$ is the onsite Hamiltonian of Cr atom, $H_{\rm X}$ is onsite Hamiltonian of ligand atoms. $T_{\rm MX}$ is the hopping matrix between M site and X site. For CrX$_3$, the 6 ligand X atoms can be related by symmetry. We only present one of the hopping matrices $T_{\rm MX_1}$, as shown in Fig. \ref{fig:illustration}(b). 
}

 {
The hopping matrix $H_{\rm MX_1}$ can be parametrized as
\begin{equation}\label{eq:distortion}
       T_{MX}= \left(
       \begin{array}{ccc}
        t_1 & \delta\tau_1 &\delta\tau_2\\
        -t_2 & \delta\tau_3 & \delta\tau_4\\
        \delta t_1 & \delta t_2 & \delta t_3\\
        \delta t_4 & \delta t_6 & t_{a0}\\
        \delta t_5 & t_{b0} & \delta t_7
    \end{array}\right).
\end{equation}
This is written in the basis of $d$ orbitals with sequence $d_{x^2}$, $d_{z^2}$, $d_{yz}$, $d_{xz}$, $d_{xy}$, and $p$ orbitals with sequence $p_x$, $p_y$, $p_z$.
The hoppings starting with $\delta$ are distortion induced hoppings and should be zero for an ideal octahedron. 
}

 {
From Wannier90 calculation, we have
\begin{equation}
       H^{\textrm{CrCl}_3}_{\rm M}= \left(
    \begin{array}{ccccc}
2262.08 & -0.02 & -11.92 & 11.61 & 0.33\\
-0.02 & 2262.48 & 6.72 & 7.27 & -13.88\\
-11.92 & 6.72 & 1710.62 & 15.46 & 15.14\\
11.61 & 7.27 & 15.46 & 1710.65 & 15.15\\
0.33 & -13.88 & 15.14 & 15.15 & 1710.82
    \end{array}\right),
\end{equation}
\begin{equation}
       H^{\textrm{CrBr}_3}_{\rm M}= \left(
    \begin{array}{ccccc}
1778.06 & 0.01 & -9.36 & 9.84 & -0.49\\
0.01 & 1778.36 & 6.1 & 5.27 & -11.29\\
-9.36 & 6.1 & 1380.58 & 9.86 & 9.66\\
9.84 & 5.27 & 9.86 & 1380.56 & 9.65\\
-0.49 & -11.29 & 9.66 & 9.65 & 1380.7
    \end{array}\right),
\end{equation}
\begin{equation}
       H^{\textrm{CrI}_3}_{\rm M}= \left(
    \begin{array}{ccccc}
3781.78 & -0.01 & -5.89 & 5.33 & 0.57\\
-0.01 & 3781.9 & 2.84 & 3.81 & -6.57\\
-5.89 & 2.84 & 3493.5 & 4.81 & 4.7\\
5.33 & 3.81 & 4.81 & 3493.51 & 4.7\\
0.57 & -6.57 & 4.7 & 4.7 & 3493.61
    \end{array}\right),
\end{equation}
\begin{equation}
       H^{\textrm{CrCl}_3}_{\rm X}= \left(
    \begin{array}{ccc}
-1140.48 & 75.56 & -15.88\\
75.56 & -1139.88 & -13.67\\
-15.88 & -13.67 & -503.56
    \end{array}\right),
\end{equation}
\begin{equation}
       H^{\textrm{CrBr}_3}_{\rm X}= \left(
    \begin{array}{ccc}
-1095.96 & 53.85 & -13.78\\
53.85 & -1094.1 & -15.86\\
-13.78 & -15.86 & -478.33
    \end{array}\right),
\end{equation}
\begin{equation}
       H^{\textrm{CrI}_3}_{\rm X}= \left(
    \begin{array}{ccc}
1415.99 & 36.19 & -11.97\\
36.19 & 1411.57 & -9.43\\
-11.97 & -9.43 & 1952.06
    \end{array}\right),
\end{equation}
\begin{equation}
       T^{\textrm{CrCl}_3}_{\rm MX}= \left(
    \begin{array}{ccc}
-1229.15 & 125.02 & -46.19\\
698.08 & -61.12 & 37.92\\
0.24 & 23.36 & -35.49\\
-66.75 & 5.03 & 725.57\\
121.75 & 718.61 & 2.26
    \end{array}\right),
\end{equation}
\begin{equation}
       T^{\textrm{CrBr}_3}_{\rm MX}= \left(
    \begin{array}{ccc}
-1098.54 & 91.48 & -39.8\\
624 & -45.81 & 31.18\\
1.7 & 19.23 & -26.56\\
-56.48 & 2.76 & 644.53\\
91.77 & 641.93 & 2.23
    \end{array}\right),
\end{equation}
\begin{equation}
       T^{\textrm{CrI}_3}_{\rm MX}= \left(
    \begin{array}{ccc}
-939.04 & 48.7 & -31.06\\
533.92 & -36.11 & 21.24\\
1.4 & 14.77 & -14.06\\
-44.83 & 1.61 & 556.07\\
41.72 & 547.73 & 0.35
    \end{array}\right).
\end{equation}
From the above parameters, we can obtain $\Delta_{pd}$, $\Delta_c^0$, $\delta^0$, $t_{pd\pi}$, $t_{pd\sigma}$ as shown in Table \ref{table:dft}, and distortion induced hoppings in Eq. (\ref{eq:distortion}).
}

\bibliography{references}

\begin{thebibliography}{44}%
\makeatletter
\providecommand \@ifxundefined [1]{%
 \@ifx{#1\undefined}
}%
\providecommand \@ifnum [1]{%
 \ifnum #1\expandafter \@firstoftwo
 \else \expandafter \@secondoftwo
 \fi
}%
\providecommand \@ifx [1]{%
 \ifx #1\expandafter \@firstoftwo
 \else \expandafter \@secondoftwo
 \fi
}%
\providecommand \natexlab [1]{#1}%
\providecommand \enquote  [1]{``#1''}%
\providecommand \bibnamefont  [1]{#1}%
\providecommand \bibfnamefont [1]{#1}%
\providecommand \citenamefont [1]{#1}%
\providecommand \href@noop [0]{\@secondoftwo}%
\providecommand \href [0]{\begingroup \@sanitize@url \@href}%
\providecommand \@href[1]{\@@startlink{#1}\@@href}%
\providecommand \@@href[1]{\endgroup#1\@@endlink}%
\providecommand \@sanitize@url [0]{\catcode `\\12\catcode `\$12\catcode
  `\&12\catcode `\#12\catcode `\^12\catcode `\_12\catcode `\%12\relax}%
\providecommand \@@startlink[1]{}%
\providecommand \@@endlink[0]{}%
\providecommand \url  [0]{\begingroup\@sanitize@url \@url }%
\providecommand \@url [1]{\endgroup\@href {#1}{\urlprefix }}%
\providecommand \urlprefix  [0]{URL }%
\providecommand \Eprint [0]{\href }%
\providecommand \doibase [0]{https://doi.org/}%
\providecommand \selectlanguage [0]{\@gobble}%
\providecommand \bibinfo  [0]{\@secondoftwo}%
\providecommand \bibfield  [0]{\@secondoftwo}%
\providecommand \translation [1]{[#1]}%
\providecommand \BibitemOpen [0]{}%
\providecommand \bibitemStop [0]{}%
\providecommand \bibitemNoStop [0]{.\EOS\space}%
\providecommand \EOS [0]{\spacefactor3000\relax}%
\providecommand \BibitemShut  [1]{\csname bibitem#1\endcsname}%
\let\auto@bib@innerbib\@empty
\bibitem [{\citenamefont {Soumyanarayanan}\ \emph {et~al.}(2016)\citenamefont
  {Soumyanarayanan}, \citenamefont {Reyren}, \citenamefont {Fert},\ and\
  \citenamefont {Panagopoulos}}]{soumyanarayanan_emergent_2016}%
  \BibitemOpen
  \bibfield  {author} {\bibinfo {author} {\bibfnamefont {A.}~\bibnamefont
  {Soumyanarayanan}}, \bibinfo {author} {\bibfnamefont {N.}~\bibnamefont
  {Reyren}}, \bibinfo {author} {\bibfnamefont {A.}~\bibnamefont {Fert}},\ and\
  \bibinfo {author} {\bibfnamefont {C.}~\bibnamefont {Panagopoulos}},\
  }\bibfield  {title} {\bibinfo {title} {Emergent phenomena induced by
  spin–orbit coupling at surfaces and interfaces},\ }\href
  {https://doi.org/10.1038/nature19820} {\bibfield  {journal} {\bibinfo
  {journal} {Nature}\ }\textbf {\bibinfo {volume} {539}},\ \bibinfo {pages}
  {509} (\bibinfo {year} {2016})}\BibitemShut {NoStop}%
\bibitem [{\citenamefont {Miao}\ \emph {et~al.}(2018)\citenamefont {Miao},
  \citenamefont {Xu}, \citenamefont {Zhu}, \citenamefont {Zhou},\ and\
  \citenamefont {Sun}}]{miao_2d_2018}%
  \BibitemOpen
  \bibfield  {author} {\bibinfo {author} {\bibfnamefont {N.}~\bibnamefont
  {Miao}}, \bibinfo {author} {\bibfnamefont {B.}~\bibnamefont {Xu}}, \bibinfo
  {author} {\bibfnamefont {L.}~\bibnamefont {Zhu}}, \bibinfo {author}
  {\bibfnamefont {J.}~\bibnamefont {Zhou}},\ and\ \bibinfo {author}
  {\bibfnamefont {Z.}~\bibnamefont {Sun}},\ }\bibfield  {title} {\bibinfo
  {title} {{2D} {Intrinsic} {Ferromagnet} from van der {Waals}
  {Antiferromagnets}},\ }\href {https://doi.org/10.1021/jacs.7b12976}
  {\bibfield  {journal} {\bibinfo  {journal} {J. Am. Chem. Soc.}\ }\textbf
  {\bibinfo {volume} {140}},\ \bibinfo {pages} {2417} (\bibinfo {year}
  {2018})}\BibitemShut {NoStop}%
\bibitem [{\citenamefont {Huang}\ \emph {et~al.}(2017)\citenamefont {Huang},
  \citenamefont {Clark}, \citenamefont {Navarro-Moratalla}, \citenamefont
  {Klein}, \citenamefont {Cheng}, \citenamefont {Seyler}, \citenamefont
  {Zhong}, \citenamefont {Schmidgall}, \citenamefont {McGuire}, \citenamefont
  {Cobden}, \citenamefont {Yao}, \citenamefont {Xiao}, \citenamefont
  {Jarillo-Herrero},\ and\ \citenamefont {Xu}}]{huang_bevin_2017}%
  \BibitemOpen
  \bibfield  {author} {\bibinfo {author} {\bibfnamefont {B.}~\bibnamefont
  {Huang}}, \bibinfo {author} {\bibfnamefont {G.}~\bibnamefont {Clark}},
  \bibinfo {author} {\bibfnamefont {E.}~\bibnamefont {Navarro-Moratalla}},
  \bibinfo {author} {\bibfnamefont {D.~R.}\ \bibnamefont {Klein}}, \bibinfo
  {author} {\bibfnamefont {R.}~\bibnamefont {Cheng}}, \bibinfo {author}
  {\bibfnamefont {K.~L.}\ \bibnamefont {Seyler}}, \bibinfo {author}
  {\bibfnamefont {D.}~\bibnamefont {Zhong}}, \bibinfo {author} {\bibfnamefont
  {E.}~\bibnamefont {Schmidgall}}, \bibinfo {author} {\bibfnamefont {M.~A.}\
  \bibnamefont {McGuire}}, \bibinfo {author} {\bibfnamefont {D.~H.}\
  \bibnamefont {Cobden}}, \bibinfo {author} {\bibfnamefont {W.}~\bibnamefont
  {Yao}}, \bibinfo {author} {\bibfnamefont {D.}~\bibnamefont {Xiao}}, \bibinfo
  {author} {\bibfnamefont {P.}~\bibnamefont {Jarillo-Herrero}},\ and\ \bibinfo
  {author} {\bibfnamefont {X.}~\bibnamefont {Xu}},\ }\bibfield  {title}
  {\bibinfo {title} {Layer-dependent ferromagnetism in a van der {Waals}
  crystal down to the monolayer limit},\ }\href
  {https://doi.org/10.1038/nature22391} {\bibfield  {journal} {\bibinfo
  {journal} {Nature}\ }\textbf {\bibinfo {volume} {546}},\ \bibinfo {pages}
  {270} (\bibinfo {year} {2017})}\BibitemShut {NoStop}%
\bibitem [{\citenamefont {Gong}\ \emph {et~al.}(2017)\citenamefont {Gong},
  \citenamefont {Li}, \citenamefont {Li}, \citenamefont {Ji}, \citenamefont
  {Stern}, \citenamefont {Xia}, \citenamefont {Cao}, \citenamefont {Bao},
  \citenamefont {Wang}, \citenamefont {Wang}, \citenamefont {Qiu},
  \citenamefont {Cava}, \citenamefont {Louie}, \citenamefont {Xia},\ and\
  \citenamefont {Zhang}}]{gong_cr2ge2te6_2017}%
  \BibitemOpen
  \bibfield  {author} {\bibinfo {author} {\bibfnamefont {C.}~\bibnamefont
  {Gong}}, \bibinfo {author} {\bibfnamefont {L.}~\bibnamefont {Li}}, \bibinfo
  {author} {\bibfnamefont {Z.}~\bibnamefont {Li}}, \bibinfo {author}
  {\bibfnamefont {H.}~\bibnamefont {Ji}}, \bibinfo {author} {\bibfnamefont
  {A.}~\bibnamefont {Stern}}, \bibinfo {author} {\bibfnamefont
  {Y.}~\bibnamefont {Xia}}, \bibinfo {author} {\bibfnamefont {T.}~\bibnamefont
  {Cao}}, \bibinfo {author} {\bibfnamefont {W.}~\bibnamefont {Bao}}, \bibinfo
  {author} {\bibfnamefont {C.}~\bibnamefont {Wang}}, \bibinfo {author}
  {\bibfnamefont {Y.}~\bibnamefont {Wang}}, \bibinfo {author} {\bibfnamefont
  {Z.~Q.}\ \bibnamefont {Qiu}}, \bibinfo {author} {\bibfnamefont {R.~J.}\
  \bibnamefont {Cava}}, \bibinfo {author} {\bibfnamefont {S.~G.}\ \bibnamefont
  {Louie}}, \bibinfo {author} {\bibfnamefont {J.}~\bibnamefont {Xia}},\ and\
  \bibinfo {author} {\bibfnamefont {X.}~\bibnamefont {Zhang}},\ }\bibfield
  {title} {\bibinfo {title} {Discovery of intrinsic ferromagnetism in
  two-dimensional van der {Waals} crystals},\ }\href
  {https://doi.org/10.1038/nature22060} {\bibfield  {journal} {\bibinfo
  {journal} {Nature}\ }\textbf {\bibinfo {volume} {546}},\ \bibinfo {pages}
  {265} (\bibinfo {year} {2017})}\BibitemShut {NoStop}%
\bibitem [{\citenamefont {Huang}\ \emph {et~al.}(2018)\citenamefont {Huang},
  \citenamefont {Clark}, \citenamefont {Klein}, \citenamefont {MacNeill},
  \citenamefont {Navarro-Moratalla}, \citenamefont {Seyler}, \citenamefont
  {Wilson}, \citenamefont {McGuire}, \citenamefont {Cobden}, \citenamefont
  {Xiao}, \citenamefont {Yao}, \citenamefont {Jarillo-Herrero},\ and\
  \citenamefont {Xu}}]{huang_gating_CrI3_2018}%
  \BibitemOpen
  \bibfield  {author} {\bibinfo {author} {\bibfnamefont {B.}~\bibnamefont
  {Huang}}, \bibinfo {author} {\bibfnamefont {G.}~\bibnamefont {Clark}},
  \bibinfo {author} {\bibfnamefont {D.~R.}\ \bibnamefont {Klein}}, \bibinfo
  {author} {\bibfnamefont {D.}~\bibnamefont {MacNeill}}, \bibinfo {author}
  {\bibfnamefont {E.}~\bibnamefont {Navarro-Moratalla}}, \bibinfo {author}
  {\bibfnamefont {K.~L.}\ \bibnamefont {Seyler}}, \bibinfo {author}
  {\bibfnamefont {N.}~\bibnamefont {Wilson}}, \bibinfo {author} {\bibfnamefont
  {M.~A.}\ \bibnamefont {McGuire}}, \bibinfo {author} {\bibfnamefont {D.~H.}\
  \bibnamefont {Cobden}}, \bibinfo {author} {\bibfnamefont {D.}~\bibnamefont
  {Xiao}}, \bibinfo {author} {\bibfnamefont {W.}~\bibnamefont {Yao}}, \bibinfo
  {author} {\bibfnamefont {P.}~\bibnamefont {Jarillo-Herrero}},\ and\ \bibinfo
  {author} {\bibfnamefont {X.}~\bibnamefont {Xu}},\ }\bibfield  {title}
  {\bibinfo {title} {Electrical control of {2D} magnetism in bilayer {CrI3}},\
  }\href {https://doi.org/10.1038/s41565-018-0121-3} {\bibfield  {journal}
  {\bibinfo  {journal} {Nat. Nanotechnol.}\ }\textbf {\bibinfo {volume} {13}},\
  \bibinfo {pages} {544} (\bibinfo {year} {2018})}\BibitemShut {NoStop}%
\bibitem [{\citenamefont {Sivadas}\ \emph {et~al.}(2018)\citenamefont
  {Sivadas}, \citenamefont {Okamoto}, \citenamefont {Xu}, \citenamefont
  {Fennie},\ and\ \citenamefont {Xiao}}]{sivadas_stacking-dependent_2018}%
  \BibitemOpen
  \bibfield  {author} {\bibinfo {author} {\bibfnamefont {N.}~\bibnamefont
  {Sivadas}}, \bibinfo {author} {\bibfnamefont {S.}~\bibnamefont {Okamoto}},
  \bibinfo {author} {\bibfnamefont {X.}~\bibnamefont {Xu}}, \bibinfo {author}
  {\bibfnamefont {C.~J.}\ \bibnamefont {Fennie}},\ and\ \bibinfo {author}
  {\bibfnamefont {D.}~\bibnamefont {Xiao}},\ }\bibfield  {title} {\bibinfo
  {title} {Stacking-{Dependent} {Magnetism} in {Bilayer} {$\mathrm{CrI}_3$}},\
  }\href {https://doi.org/10.1021/acs.nanolett.8b03321} {\bibfield  {journal}
  {\bibinfo  {journal} {Nano Lett.}\ }\textbf {\bibinfo {volume} {18}},\
  \bibinfo {pages} {7658} (\bibinfo {year} {2018})}\BibitemShut {NoStop}%
\bibitem [{\citenamefont {Webster}\ and\ \citenamefont
  {Yan}(2018)}]{webster_strain_CrX3_2018}%
  \BibitemOpen
  \bibfield  {author} {\bibinfo {author} {\bibfnamefont {L.}~\bibnamefont
  {Webster}}\ and\ \bibinfo {author} {\bibfnamefont {J.-A.}\ \bibnamefont
  {Yan}},\ }\bibfield  {title} {\bibinfo {title} {Strain-tunable magnetic
  anisotropy in monolayer {${\mathrm{CrCl}}_{3}$}, {${\mathrm{CrBr}}_{3}$}, and
  {${\mathrm{CrI}}_{3}$}},\ }\href {https://doi.org/10.1103/PhysRevB.98.144411}
  {\bibfield  {journal} {\bibinfo  {journal} {Phys. Rev. B}\ }\textbf {\bibinfo
  {volume} {98}},\ \bibinfo {pages} {144411} (\bibinfo {year}
  {2018})}\BibitemShut {NoStop}%
\bibitem [{\citenamefont {Wu}\ \emph {et~al.}(2019)\citenamefont {Wu},
  \citenamefont {Yu},\ and\ \citenamefont
  {Yuan}}]{wu_strain-tunable_CrI3_2019}%
  \BibitemOpen
  \bibfield  {author} {\bibinfo {author} {\bibfnamefont {Z.}~\bibnamefont
  {Wu}}, \bibinfo {author} {\bibfnamefont {J.}~\bibnamefont {Yu}},\ and\
  \bibinfo {author} {\bibfnamefont {S.}~\bibnamefont {Yuan}},\ }\bibfield
  {title} {\bibinfo {title} {Strain-tunable magnetic and electronic properties
  of monolayer {$\mathrm{CrI}_3$}},\ }\href
  {https://doi.org/10.1039/C8CP07067A} {\bibfield  {journal} {\bibinfo
  {journal} {Phys. Chem. Chem. Phys.}\ }\textbf {\bibinfo {volume} {21}},\
  \bibinfo {pages} {7750} (\bibinfo {year} {2019})}\BibitemShut {NoStop}%
\bibitem [{\citenamefont {Cai}\ \emph {et~al.}(2019)\citenamefont {Cai},
  \citenamefont {Song}, \citenamefont {Wilson}, \citenamefont {Clark},
  \citenamefont {He}, \citenamefont {Zhang}, \citenamefont {Taniguchi},
  \citenamefont {Watanabe}, \citenamefont {Yao}, \citenamefont {Xiao},
  \citenamefont {McGuire}, \citenamefont {Cobden},\ and\ \citenamefont
  {Xu}}]{cai_atomically_crcl3_2019}%
  \BibitemOpen
  \bibfield  {author} {\bibinfo {author} {\bibfnamefont {X.}~\bibnamefont
  {Cai}}, \bibinfo {author} {\bibfnamefont {T.}~\bibnamefont {Song}}, \bibinfo
  {author} {\bibfnamefont {N.~P.}\ \bibnamefont {Wilson}}, \bibinfo {author}
  {\bibfnamefont {G.}~\bibnamefont {Clark}}, \bibinfo {author} {\bibfnamefont
  {M.}~\bibnamefont {He}}, \bibinfo {author} {\bibfnamefont {X.}~\bibnamefont
  {Zhang}}, \bibinfo {author} {\bibfnamefont {T.}~\bibnamefont {Taniguchi}},
  \bibinfo {author} {\bibfnamefont {K.}~\bibnamefont {Watanabe}}, \bibinfo
  {author} {\bibfnamefont {W.}~\bibnamefont {Yao}}, \bibinfo {author}
  {\bibfnamefont {D.}~\bibnamefont {Xiao}}, \bibinfo {author} {\bibfnamefont
  {M.~A.}\ \bibnamefont {McGuire}}, \bibinfo {author} {\bibfnamefont {D.~H.}\
  \bibnamefont {Cobden}},\ and\ \bibinfo {author} {\bibfnamefont
  {X.}~\bibnamefont {Xu}},\ }\bibfield  {title} {\bibinfo {title} {Atomically
  {Thin} {$\mathrm{CrCl}_3$}: {An} {In}-{Plane} {Layered} {Antiferromagnetic}
  {Insulator}},\ }\href {https://doi.org/10.1021/acs.nanolett.9b01317}
  {\bibfield  {journal} {\bibinfo  {journal} {Nano Lett.}\ }\textbf {\bibinfo
  {volume} {19}},\ \bibinfo {pages} {3993} (\bibinfo {year}
  {2019})}\BibitemShut {NoStop}%
\bibitem [{\citenamefont {Li}\ \emph {et~al.}(2019)\citenamefont {Li},
  \citenamefont {Jiang}, \citenamefont {Sivadas}, \citenamefont {Wang},
  \citenamefont {Xu}, \citenamefont {Weber}, \citenamefont {Goldberger},
  \citenamefont {Watanabe}, \citenamefont {Taniguchi}, \citenamefont {Fennie},
  \citenamefont {Fai~Mak},\ and\ \citenamefont {Shan}}]{li_pressure_CrI3_2019}%
  \BibitemOpen
  \bibfield  {author} {\bibinfo {author} {\bibfnamefont {T.}~\bibnamefont
  {Li}}, \bibinfo {author} {\bibfnamefont {S.}~\bibnamefont {Jiang}}, \bibinfo
  {author} {\bibfnamefont {N.}~\bibnamefont {Sivadas}}, \bibinfo {author}
  {\bibfnamefont {Z.}~\bibnamefont {Wang}}, \bibinfo {author} {\bibfnamefont
  {Y.}~\bibnamefont {Xu}}, \bibinfo {author} {\bibfnamefont {D.}~\bibnamefont
  {Weber}}, \bibinfo {author} {\bibfnamefont {J.~E.}\ \bibnamefont
  {Goldberger}}, \bibinfo {author} {\bibfnamefont {K.}~\bibnamefont
  {Watanabe}}, \bibinfo {author} {\bibfnamefont {T.}~\bibnamefont {Taniguchi}},
  \bibinfo {author} {\bibfnamefont {C.~J.}\ \bibnamefont {Fennie}}, \bibinfo
  {author} {\bibfnamefont {K.}~\bibnamefont {Fai~Mak}},\ and\ \bibinfo {author}
  {\bibfnamefont {J.}~\bibnamefont {Shan}},\ }\bibfield  {title} {\bibinfo
  {title} {Pressure-controlled interlayer magnetism in atomically thin
  {CrI$_3$}},\ }\href {https://doi.org/10.1038/s41563-019-0506-1} {\bibfield
  {journal} {\bibinfo  {journal} {Nat. Mater.}\ }\textbf {\bibinfo {volume}
  {18}},\ \bibinfo {pages} {1303} (\bibinfo {year} {2019})}\BibitemShut
  {NoStop}%
\bibitem [{\citenamefont {Mermin}\ and\ \citenamefont
  {Wagner}(1966)}]{mermin_wagner_1966}%
  \BibitemOpen
  \bibfield  {author} {\bibinfo {author} {\bibfnamefont {N.~D.}\ \bibnamefont
  {Mermin}}\ and\ \bibinfo {author} {\bibfnamefont {H.}~\bibnamefont
  {Wagner}},\ }\bibfield  {title} {\bibinfo {title} {Absence of
  {Ferromagnetism} or {Antiferromagnetism} in {One}- or {Two}-{Dimensional}
  {Isotropic} {Heisenberg} {Models}},\ }\href
  {https://doi.org/10.1103/PhysRevLett.17.1133} {\bibfield  {journal} {\bibinfo
   {journal} {Phys. Rev. Lett.}\ }\textbf {\bibinfo {volume} {17}},\ \bibinfo
  {pages} {1133} (\bibinfo {year} {1966})}\BibitemShut {NoStop}%
\bibitem [{\citenamefont {Lado}\ and\ \citenamefont
  {Fernández-Rossier}(2017)}]{lado_XXZ_2017}%
  \BibitemOpen
  \bibfield  {author} {\bibinfo {author} {\bibfnamefont {J.~L.}\ \bibnamefont
  {Lado}}\ and\ \bibinfo {author} {\bibfnamefont {J.}~\bibnamefont
  {Fernández-Rossier}},\ }\bibfield  {title} {\bibinfo {title} {On the origin
  of magnetic anisotropy in two dimensional $\mathrm{CrI}_3$},\ }\href
  {https://doi.org/10.1088/2053-1583/aa75ed} {\bibfield  {journal} {\bibinfo
  {journal} {2D Materials}\ }\textbf {\bibinfo {volume} {4}},\ \bibinfo {pages}
  {035002} (\bibinfo {year} {2017})}\BibitemShut {NoStop}%
\bibitem [{\citenamefont {Kim}\ \emph {et~al.}(2019)\citenamefont {Kim},
  \citenamefont {Kim}, \citenamefont {Ko}, \citenamefont {Seo}, \citenamefont
  {Kim}, \citenamefont {Jang}, \citenamefont {Kim}, \citenamefont {Kim},
  \citenamefont {Cheong},\ and\ \citenamefont {Park}}]{kim_XXZ_2019}%
  \BibitemOpen
  \bibfield  {author} {\bibinfo {author} {\bibfnamefont {D.-H.}\ \bibnamefont
  {Kim}}, \bibinfo {author} {\bibfnamefont {K.}~\bibnamefont {Kim}}, \bibinfo
  {author} {\bibfnamefont {K.-T.}\ \bibnamefont {Ko}}, \bibinfo {author}
  {\bibfnamefont {J.}~\bibnamefont {Seo}}, \bibinfo {author} {\bibfnamefont
  {J.~S.}\ \bibnamefont {Kim}}, \bibinfo {author} {\bibfnamefont {T.-H.}\
  \bibnamefont {Jang}}, \bibinfo {author} {\bibfnamefont {Y.}~\bibnamefont
  {Kim}}, \bibinfo {author} {\bibfnamefont {J.-Y.}\ \bibnamefont {Kim}},
  \bibinfo {author} {\bibfnamefont {S.-W.}\ \bibnamefont {Cheong}},\ and\
  \bibinfo {author} {\bibfnamefont {J.-H.}\ \bibnamefont {Park}},\ }\bibfield
  {title} {\bibinfo {title} {Giant magnetic anisotropy induced by ligand {LS}
  coupling in layered {Cr} compounds},\ }\href
  {https://doi.org/10.1103/PhysRevLett.122.207201} {\bibfield  {journal}
  {\bibinfo  {journal} {Phys. Rev. Lett.}\ }\textbf {\bibinfo {volume} {122}},\
  \bibinfo {pages} {207201} (\bibinfo {year} {2019})}\BibitemShut {NoStop}%
\bibitem [{\citenamefont {Xu}\ \emph {et~al.}(2018)\citenamefont {Xu},
  \citenamefont {Feng}, \citenamefont {Xiang},\ and\ \citenamefont
  {Bellaiche}}]{xu_SIAK_2018}%
  \BibitemOpen
  \bibfield  {author} {\bibinfo {author} {\bibfnamefont {C.}~\bibnamefont
  {Xu}}, \bibinfo {author} {\bibfnamefont {J.}~\bibnamefont {Feng}}, \bibinfo
  {author} {\bibfnamefont {H.}~\bibnamefont {Xiang}},\ and\ \bibinfo {author}
  {\bibfnamefont {L.}~\bibnamefont {Bellaiche}},\ }\bibfield  {title} {\bibinfo
  {title} {Interplay between {Kitaev} interaction and single ion anisotropy in
  ferromagnetic $\mathrm{CrI}_3$ and $\mathrm{CrGeTe}_3$ monolayers},\ }\href
  {https://doi.org/10.1038/s41524-018-0115-6} {\bibfield  {journal} {\bibinfo
  {journal} {Npj Comput. Mater.}\ }\textbf {\bibinfo {volume} {4}},\ \bibinfo
  {pages} {1} (\bibinfo {year} {2018})}\BibitemShut {NoStop}%
\bibitem [{\citenamefont {Lee}\ \emph {et~al.}(2020)\citenamefont {Lee},
  \citenamefont {Utermohlen}, \citenamefont {Weber}, \citenamefont {Hwang},
  \citenamefont {Zhang}, \citenamefont {van Tol}, \citenamefont {Goldberger},
  \citenamefont {Trivedi},\ and\ \citenamefont {Hammel}}]{lee_KGammal_2020}%
  \BibitemOpen
  \bibfield  {author} {\bibinfo {author} {\bibfnamefont {I.}~\bibnamefont
  {Lee}}, \bibinfo {author} {\bibfnamefont {F.~G.}\ \bibnamefont {Utermohlen}},
  \bibinfo {author} {\bibfnamefont {D.}~\bibnamefont {Weber}}, \bibinfo
  {author} {\bibfnamefont {K.}~\bibnamefont {Hwang}}, \bibinfo {author}
  {\bibfnamefont {C.}~\bibnamefont {Zhang}}, \bibinfo {author} {\bibfnamefont
  {J.}~\bibnamefont {van Tol}}, \bibinfo {author} {\bibfnamefont {J.~E.}\
  \bibnamefont {Goldberger}}, \bibinfo {author} {\bibfnamefont
  {N.}~\bibnamefont {Trivedi}},\ and\ \bibinfo {author} {\bibfnamefont {P.~C.}\
  \bibnamefont {Hammel}},\ }\bibfield  {title} {\bibinfo {title} {{Fundamental}
  {Spin} {Interactions} {Underlying} the {Magnetic} {Anisotropy} in the
  {Kitaev} {Ferromagnet} $\mathrm{CrI}_3$},\ }\href
  {https://doi.org/10.1103/PhysRevLett.124.017201} {\bibfield  {journal}
  {\bibinfo  {journal} {Phys. Rev. Lett.}\ }\textbf {\bibinfo {volume} {124}},\
  \bibinfo {pages} {017201} (\bibinfo {year} {2020})}\BibitemShut {NoStop}%
\bibitem [{\citenamefont {Kanamori}(1963)}]{kanamori_1963}%
  \BibitemOpen
  \bibfield  {author} {\bibinfo {author} {\bibfnamefont {J.}~\bibnamefont
  {Kanamori}},\ }\bibfield  {title} {\bibinfo {title} {Electron correlation and
  ferromagnetism of transition metals},\ }\href
  {https://doi.org/10.1143/PTP.30.275} {\bibfield  {journal} {\bibinfo
  {journal} {Prog. Theor. Phys.}\ }\textbf {\bibinfo {volume} {30}},\ \bibinfo
  {pages} {275} (\bibinfo {year} {1963})}\BibitemShut {NoStop}%
\bibitem [{\citenamefont {Jackeli}\ and\ \citenamefont
  {Khaliullin}(2009)}]{jackeli_mott_2009}%
  \BibitemOpen
  \bibfield  {author} {\bibinfo {author} {\bibfnamefont {G.}~\bibnamefont
  {Jackeli}}\ and\ \bibinfo {author} {\bibfnamefont {G.}~\bibnamefont
  {Khaliullin}},\ }\bibfield  {title} {\bibinfo {title} {{Mott} {Insulators} in
  the {Strong} {Spin}-{Orbit} {Coupling} {Limit}: {From} {Heisenberg} to a
  {Quantum} {Compass} and {Kitaev} {Models}},\ }\href
  {https://doi.org/10.1103/PhysRevLett.102.017205} {\bibfield  {journal}
  {\bibinfo  {journal} {Phys. Rev. Lett.}\ }\textbf {\bibinfo {volume} {102}},\
  \bibinfo {pages} {017205} (\bibinfo {year} {2009})}\BibitemShut {NoStop}%
\bibitem [{\citenamefont {Rau}\ \emph {et~al.}(2014)\citenamefont {Rau},
  \citenamefont {Lee},\ and\ \citenamefont {Kee}}]{rau_spinmodel_2014}%
  \BibitemOpen
  \bibfield  {author} {\bibinfo {author} {\bibfnamefont {J.~G.}\ \bibnamefont
  {Rau}}, \bibinfo {author} {\bibfnamefont {E.~K.-H.}\ \bibnamefont {Lee}},\
  and\ \bibinfo {author} {\bibfnamefont {H.-Y.}\ \bibnamefont {Kee}},\
  }\bibfield  {title} {\bibinfo {title} {Generic {Spin} {Model} for the
  {Honeycomb} {Iridates} beyond the {Kitaev} {Limit}},\ }\href
  {https://doi.org/10.1103/PhysRevLett.112.077204} {\bibfield  {journal}
  {\bibinfo  {journal} {Phys. Rev. Lett.}\ }\textbf {\bibinfo {volume} {112}},\
  \bibinfo {pages} {077204} (\bibinfo {year} {2014})}\BibitemShut {NoStop}%
\bibitem [{\citenamefont {Kim}\ \emph {et~al.}(2015)\citenamefont {Kim},
  \citenamefont {V.}, \citenamefont {Catuneanu},\ and\ \citenamefont
  {Kee}}]{kim_rucl3_2015}%
  \BibitemOpen
  \bibfield  {author} {\bibinfo {author} {\bibfnamefont {H.-S.}\ \bibnamefont
  {Kim}}, \bibinfo {author} {\bibfnamefont {V.~S.}\ \bibnamefont {V.}},
  \bibinfo {author} {\bibfnamefont {A.}~\bibnamefont {Catuneanu}},\ and\
  \bibinfo {author} {\bibfnamefont {H.-Y.}\ \bibnamefont {Kee}},\ }\bibfield
  {title} {\bibinfo {title} {Kitaev magnetism in honeycomb
  {${\text{RuCl}}_{3}$} with intermediate spin-orbit coupling},\ }\href
  {https://doi.org/10.1103/PhysRevB.91.241110} {\bibfield  {journal} {\bibinfo
  {journal} {Phys. Rev. B}\ }\textbf {\bibinfo {volume} {91}},\ \bibinfo
  {pages} {241110} (\bibinfo {year} {2015})}\BibitemShut {NoStop}%
\bibitem [{\citenamefont {Stavropoulos}\ \emph {et~al.}(2019)\citenamefont
  {Stavropoulos}, \citenamefont {Pereira},\ and\ \citenamefont
  {Kee}}]{stavropoulos_S=1_2019}%
  \BibitemOpen
  \bibfield  {author} {\bibinfo {author} {\bibfnamefont {P.~P.}\ \bibnamefont
  {Stavropoulos}}, \bibinfo {author} {\bibfnamefont {D.}~\bibnamefont
  {Pereira}},\ and\ \bibinfo {author} {\bibfnamefont {H.-Y.}\ \bibnamefont
  {Kee}},\ }\bibfield  {title} {\bibinfo {title} {Microscopic {Mechanism} for a
  {Higher}-{Spin} {Kitaev} {Model}},\ }\href
  {https://doi.org/10.1103/PhysRevLett.123.037203} {\bibfield  {journal}
  {\bibinfo  {journal} {Phys. Rev. Lett.}\ }\textbf {\bibinfo {volume} {123}},\
  \bibinfo {pages} {037203} (\bibinfo {year} {2019})}\BibitemShut {NoStop}%
\bibitem [{\citenamefont {Stavropoulos}\ \emph {et~al.}(2021)\citenamefont
  {Stavropoulos}, \citenamefont {Liu},\ and\ \citenamefont
  {Kee}}]{stavropoulos_2021}%
  \BibitemOpen
  \bibfield  {author} {\bibinfo {author} {\bibfnamefont {P.~P.}\ \bibnamefont
  {Stavropoulos}}, \bibinfo {author} {\bibfnamefont {X.}~\bibnamefont {Liu}},\
  and\ \bibinfo {author} {\bibfnamefont {H.-Y.}\ \bibnamefont {Kee}},\
  }\bibfield  {title} {\bibinfo {title} {Magnetic anisotropy in spin-3/2 with
  heavy ligand in honeycomb {Mott} insulators: {Application} to {CrI$_3$}},\
  }\href {https://doi.org/10.1103/PhysRevResearch.3.013216} {\bibfield
  {journal} {\bibinfo  {journal} {Phys. Rev. Res.}\ }\textbf {\bibinfo {volume}
  {3}},\ \bibinfo {pages} {013216} (\bibinfo {year} {2021})}\BibitemShut
  {NoStop}%
\bibitem [{\citenamefont {Liu}\ \emph {et~al.}(2020)\citenamefont {Liu},
  \citenamefont {Chaloupka},\ and\ \citenamefont
  {Khaliullin}}]{huimei_prl_2020}%
  \BibitemOpen
  \bibfield  {author} {\bibinfo {author} {\bibfnamefont {H.}~\bibnamefont
  {Liu}}, \bibinfo {author} {\bibfnamefont {J.}~\bibnamefont {Chaloupka}},\
  and\ \bibinfo {author} {\bibfnamefont {G.}~\bibnamefont {Khaliullin}},\
  }\bibfield  {title} {\bibinfo {title} {Kitaev {Spin} {Liquid} in {3D}
  {Transition} {Metal} {Compounds}},\ }\href
  {https://doi.org/10.1103/PhysRevLett.125.047201} {\bibfield  {journal}
  {\bibinfo  {journal} {Phys. Rev. Lett.}\ }\textbf {\bibinfo {volume} {125}},\
  \bibinfo {pages} {047201} (\bibinfo {year} {2020})}\BibitemShut {NoStop}%
\bibitem [{\citenamefont {Sugano}(2014)}]{sugano_multiplets_2014}%
  \BibitemOpen
  \bibfield  {author} {\bibinfo {author} {\bibfnamefont {S.}~\bibnamefont
  {Sugano}},\ }\href {http://qut.eblib.com.au/patron/FullRecord.aspx?p=1172180}
  {\emph {\bibinfo {title} {Multiplets of {Transition}-{Metal} {Ions} in
  {Crystals}.}}}\ (\bibinfo  {publisher} {Elsevier Science},\ \bibinfo
  {address} {Saint Louis},\ \bibinfo {year} {2014})\BibitemShut {NoStop}%
\bibitem [{\citenamefont {Coury}\ \emph {et~al.}(2016)\citenamefont {Coury},
  \citenamefont {Dudarev}, \citenamefont {Foulkes}, \citenamefont {Horsfield},
  \citenamefont {Ma},\ and\ \citenamefont {Spencer}}]{coury_hubbard_2016}%
  \BibitemOpen
  \bibfield  {author} {\bibinfo {author} {\bibfnamefont {M.~E.~A.}\
  \bibnamefont {Coury}}, \bibinfo {author} {\bibfnamefont {S.~L.}\ \bibnamefont
  {Dudarev}}, \bibinfo {author} {\bibfnamefont {W.~M.~C.}\ \bibnamefont
  {Foulkes}}, \bibinfo {author} {\bibfnamefont {A.~P.}\ \bibnamefont
  {Horsfield}}, \bibinfo {author} {\bibfnamefont {P.-W.}\ \bibnamefont {Ma}},\
  and\ \bibinfo {author} {\bibfnamefont {J.~S.}\ \bibnamefont {Spencer}},\
  }\bibfield  {title} {\bibinfo {title} {Hubbard-like {Hamiltonians} for
  interacting electrons in s, p, and d orbitals},\ }\bibfield  {journal}
  {\bibinfo  {journal} {Phys. Rev. B}\ }\textbf {\bibinfo {volume} {93}},\
  \href {https://doi.org/10.1103/PhysRevB.93.075101}
  {10.1103/PhysRevB.93.075101} (\bibinfo {year} {2016})\BibitemShut {NoStop}%
\bibitem [{\citenamefont {Wang}\ \emph {et~al.}(2019)\citenamefont {Wang},
  \citenamefont {Fabbris}, \citenamefont {Dean},\ and\ \citenamefont
  {Kotliar}}]{wang_edrixs_2019}%
  \BibitemOpen
  \bibfield  {author} {\bibinfo {author} {\bibfnamefont {Y.}~\bibnamefont
  {Wang}}, \bibinfo {author} {\bibfnamefont {G.}~\bibnamefont {Fabbris}},
  \bibinfo {author} {\bibfnamefont {M.}~\bibnamefont {Dean}},\ and\ \bibinfo
  {author} {\bibfnamefont {G.}~\bibnamefont {Kotliar}},\ }\bibfield  {title}
  {\bibinfo {title} {Edrixs: An open source toolkit for simulating spectra of
  resonant inelastic x-ray scattering},\ }\href
  {https://doi.org/https://doi.org/10.1016/j.cpc.2019.04.018} {\bibfield
  {journal} {\bibinfo  {journal} {Comput. Phys. Commun.}\ }\textbf {\bibinfo
  {volume} {243}},\ \bibinfo {pages} {151} (\bibinfo {year}
  {2019})}\BibitemShut {NoStop}%
\bibitem [{\citenamefont {Suzuki}\ \emph {et~al.}(2019)\citenamefont {Suzuki},
  \citenamefont {Gretarsson}, \citenamefont {Ishikawa}, \citenamefont {Ueda},
  \citenamefont {Yang}, \citenamefont {Liu}, \citenamefont {Kim}, \citenamefont
  {Kukusta}, \citenamefont {Yaresko}, \citenamefont {Minola}, \citenamefont
  {Sears}, \citenamefont {Francoual}, \citenamefont {Wille}, \citenamefont
  {Nuss}, \citenamefont {Takagi}, \citenamefont {Kim}, \citenamefont
  {Khaliullin}, \citenamefont {Yavaş},\ and\ \citenamefont
  {Keimer}}]{suzuki_spin_2019}%
  \BibitemOpen
  \bibfield  {author} {\bibinfo {author} {\bibfnamefont {H.}~\bibnamefont
  {Suzuki}}, \bibinfo {author} {\bibfnamefont {H.}~\bibnamefont {Gretarsson}},
  \bibinfo {author} {\bibfnamefont {H.}~\bibnamefont {Ishikawa}}, \bibinfo
  {author} {\bibfnamefont {K.}~\bibnamefont {Ueda}}, \bibinfo {author}
  {\bibfnamefont {Z.}~\bibnamefont {Yang}}, \bibinfo {author} {\bibfnamefont
  {H.}~\bibnamefont {Liu}}, \bibinfo {author} {\bibfnamefont {H.}~\bibnamefont
  {Kim}}, \bibinfo {author} {\bibfnamefont {D.}~\bibnamefont {Kukusta}},
  \bibinfo {author} {\bibfnamefont {A.}~\bibnamefont {Yaresko}}, \bibinfo
  {author} {\bibfnamefont {M.}~\bibnamefont {Minola}}, \bibinfo {author}
  {\bibfnamefont {J.~A.}\ \bibnamefont {Sears}}, \bibinfo {author}
  {\bibfnamefont {S.}~\bibnamefont {Francoual}}, \bibinfo {author}
  {\bibfnamefont {H.-C.}\ \bibnamefont {Wille}}, \bibinfo {author}
  {\bibfnamefont {J.}~\bibnamefont {Nuss}}, \bibinfo {author} {\bibfnamefont
  {H.}~\bibnamefont {Takagi}}, \bibinfo {author} {\bibfnamefont {B.~J.}\
  \bibnamefont {Kim}}, \bibinfo {author} {\bibfnamefont {G.}~\bibnamefont
  {Khaliullin}}, \bibinfo {author} {\bibfnamefont {H.}~\bibnamefont {Yavaş}},\
  and\ \bibinfo {author} {\bibfnamefont {B.}~\bibnamefont {Keimer}},\
  }\bibfield  {title} {\bibinfo {title} {Spin waves and spin-state transitions
  in a ruthenate high-temperature antiferromagnet},\ }\href
  {https://doi.org/10.1038/s41563-019-0327-2} {\bibfield  {journal} {\bibinfo
  {journal} {Nat. Mater.}\ }\textbf {\bibinfo {volume} {18}},\ \bibinfo {pages}
  {563} (\bibinfo {year} {2019})}\BibitemShut {NoStop}%
\bibitem [{\citenamefont {Fazekas}(1999)}]{fazekas_1999}%
  \BibitemOpen
  \bibfield  {author} {\bibinfo {author} {\bibfnamefont {P.}~\bibnamefont
  {Fazekas}},\ }\href {https://doi.org/10.1142/2945} {\emph {\bibinfo {title}
  {Lecture Notes on Electron Correlation and Magnetism}}}\ (\bibinfo {year}
  {1999})\BibitemShut {NoStop}%
\bibitem [{\citenamefont {Kresse}\ and\ \citenamefont
  {Hafner}(1993)}]{vasp1993}%
  \BibitemOpen
  \bibfield  {author} {\bibinfo {author} {\bibfnamefont {G.}~\bibnamefont
  {Kresse}}\ and\ \bibinfo {author} {\bibfnamefont {J.}~\bibnamefont
  {Hafner}},\ }\bibfield  {title} {\bibinfo {title} {\emph{Ab} \emph{initio}
  molecular dynamics for liquid metals},\ }\href
  {https://doi.org/10.1103/PhysRevB.47.558} {\bibfield  {journal} {\bibinfo
  {journal} {Phys. Rev. B}\ }\textbf {\bibinfo {volume} {47}},\ \bibinfo
  {pages} {558} (\bibinfo {year} {1993})}\BibitemShut {NoStop}%
\bibitem [{\citenamefont {Bl\"ochl}(1994)}]{paw1994}%
  \BibitemOpen
  \bibfield  {author} {\bibinfo {author} {\bibfnamefont {P.~E.}\ \bibnamefont
  {Bl\"ochl}},\ }\bibfield  {title} {\bibinfo {title} {Projector augmented-wave
  method},\ }\href {https://doi.org/10.1103/PhysRevB.50.17953} {\bibfield
  {journal} {\bibinfo  {journal} {Phys. Rev. B}\ }\textbf {\bibinfo {volume}
  {50}},\ \bibinfo {pages} {17953} (\bibinfo {year} {1994})}\BibitemShut
  {NoStop}%
\bibitem [{\citenamefont {Perdew}\ \emph {et~al.}(1996)\citenamefont {Perdew},
  \citenamefont {Burke},\ and\ \citenamefont {Ernzerhof}}]{pbe1996}%
  \BibitemOpen
  \bibfield  {author} {\bibinfo {author} {\bibfnamefont {J.~P.}\ \bibnamefont
  {Perdew}}, \bibinfo {author} {\bibfnamefont {K.}~\bibnamefont {Burke}},\ and\
  \bibinfo {author} {\bibfnamefont {M.}~\bibnamefont {Ernzerhof}},\ }\bibfield
  {title} {\bibinfo {title} {Generalized gradient approximation made simple},\
  }\href {https://doi.org/10.1103/PhysRevLett.77.3865} {\bibfield  {journal}
  {\bibinfo  {journal} {Phys. Rev. Lett.}\ }\textbf {\bibinfo {volume} {77}},\
  \bibinfo {pages} {3865} (\bibinfo {year} {1996})}\BibitemShut {NoStop}%
\bibitem [{\citenamefont {Morosin}\ and\ \citenamefont
  {Narath}(1964)}]{morosin_xray_1964}%
  \BibitemOpen
  \bibfield  {author} {\bibinfo {author} {\bibfnamefont {B.}~\bibnamefont
  {Morosin}}\ and\ \bibinfo {author} {\bibfnamefont {A.}~\bibnamefont
  {Narath}},\ }\bibfield  {title} {\bibinfo {title} {X‐{Ray} {Diffraction}
  and {Nuclear} {Quadrupole} {Resonance} {Studies} of {Chromium}
  {Trichloride}},\ }\href {https://doi.org/10.1063/1.1725428} {\bibfield
  {journal} {\bibinfo  {journal} {J. Chem. Phys.}\ }\textbf {\bibinfo {volume}
  {40}},\ \bibinfo {pages} {1958} (\bibinfo {year} {1964})}\BibitemShut
  {NoStop}%
\bibitem [{\citenamefont {Braekken}(1932)}]{braekken_kristallstruktur_1932}%
  \BibitemOpen
  \bibfield  {author} {\bibinfo {author} {\bibfnamefont {H.}~\bibnamefont
  {Braekken}},\ }\bibfield  {title} {\bibinfo {title} {Die {Kristallstruktur}
  von {Chromtribromid}},\ }\href@noop {} {\bibfield  {journal} {\bibinfo
  {journal} {Kongelige Norske Videnskapers Selskab, Forhandlinger}\ }\textbf
  {\bibinfo {volume} {5}},\ \bibinfo {pages} {42} (\bibinfo {year}
  {1932})}\BibitemShut {NoStop}%
\bibitem [{\citenamefont {McGuire}\ \emph {et~al.}(2015)\citenamefont
  {McGuire}, \citenamefont {Dixit}, \citenamefont {Cooper},\ and\ \citenamefont
  {Sales}}]{mcguire_CrI3_2015}%
  \BibitemOpen
  \bibfield  {author} {\bibinfo {author} {\bibfnamefont {M.~A.}\ \bibnamefont
  {McGuire}}, \bibinfo {author} {\bibfnamefont {H.}~\bibnamefont {Dixit}},
  \bibinfo {author} {\bibfnamefont {V.~R.}\ \bibnamefont {Cooper}},\ and\
  \bibinfo {author} {\bibfnamefont {B.~C.}\ \bibnamefont {Sales}},\ }\bibfield
  {title} {\bibinfo {title} {Coupling of {Crystal} {Structure} and {Magnetism}
  in the {Layered}, {Ferromagnetic} {Insulator} {CrI$_3$}},\ }\href
  {https://doi.org/10.1021/cm504242t} {\bibfield  {journal} {\bibinfo
  {journal} {Chem. Mater.}\ }\textbf {\bibinfo {volume} {27}},\ \bibinfo
  {pages} {612} (\bibinfo {year} {2015})}\BibitemShut {NoStop}%
\bibitem [{\citenamefont {Pizzi}\ \emph {et~al.}(2020)\citenamefont {Pizzi},
  \citenamefont {Vitale}, \citenamefont {Arita}, \citenamefont {Blügel},
  \citenamefont {Freimuth}, \citenamefont {G{\'{e}}ranton}, \citenamefont
  {Gibertini}, \citenamefont {Gresch}, \citenamefont {Johnson}, \citenamefont
  {Koretsune}, \citenamefont {Iba{\~{n}}ez-Azpiroz}, \citenamefont {Lee},
  \citenamefont {Lihm}, \citenamefont {Marchand}, \citenamefont {Marrazzo},
  \citenamefont {Mokrousov}, \citenamefont {Mustafa}, \citenamefont {Nohara},
  \citenamefont {Nomura}, \citenamefont {Paulatto}, \citenamefont
  {Ponc{\'{e}}}, \citenamefont {Ponweiser}, \citenamefont {Qiao}, \citenamefont
  {Thöle}, \citenamefont {Tsirkin}, \citenamefont {Wierzbowska}, \citenamefont
  {Marzari}, \citenamefont {Vanderbilt}, \citenamefont {Souza}, \citenamefont
  {Mostofi},\ and\ \citenamefont {Yates}}]{wannier90_2020}%
  \BibitemOpen
  \bibfield  {author} {\bibinfo {author} {\bibfnamefont {G.}~\bibnamefont
  {Pizzi}}, \bibinfo {author} {\bibfnamefont {V.}~\bibnamefont {Vitale}},
  \bibinfo {author} {\bibfnamefont {R.}~\bibnamefont {Arita}}, \bibinfo
  {author} {\bibfnamefont {S.}~\bibnamefont {Blügel}}, \bibinfo {author}
  {\bibfnamefont {F.}~\bibnamefont {Freimuth}}, \bibinfo {author}
  {\bibfnamefont {G.}~\bibnamefont {G{\'{e}}ranton}}, \bibinfo {author}
  {\bibfnamefont {M.}~\bibnamefont {Gibertini}}, \bibinfo {author}
  {\bibfnamefont {D.}~\bibnamefont {Gresch}}, \bibinfo {author} {\bibfnamefont
  {C.}~\bibnamefont {Johnson}}, \bibinfo {author} {\bibfnamefont
  {T.}~\bibnamefont {Koretsune}}, \bibinfo {author} {\bibfnamefont
  {J.}~\bibnamefont {Iba{\~{n}}ez-Azpiroz}}, \bibinfo {author} {\bibfnamefont
  {H.}~\bibnamefont {Lee}}, \bibinfo {author} {\bibfnamefont {J.-M.}\
  \bibnamefont {Lihm}}, \bibinfo {author} {\bibfnamefont {D.}~\bibnamefont
  {Marchand}}, \bibinfo {author} {\bibfnamefont {A.}~\bibnamefont {Marrazzo}},
  \bibinfo {author} {\bibfnamefont {Y.}~\bibnamefont {Mokrousov}}, \bibinfo
  {author} {\bibfnamefont {J.~I.}\ \bibnamefont {Mustafa}}, \bibinfo {author}
  {\bibfnamefont {Y.}~\bibnamefont {Nohara}}, \bibinfo {author} {\bibfnamefont
  {Y.}~\bibnamefont {Nomura}}, \bibinfo {author} {\bibfnamefont
  {L.}~\bibnamefont {Paulatto}}, \bibinfo {author} {\bibfnamefont
  {S.}~\bibnamefont {Ponc{\'{e}}}}, \bibinfo {author} {\bibfnamefont
  {T.}~\bibnamefont {Ponweiser}}, \bibinfo {author} {\bibfnamefont
  {J.}~\bibnamefont {Qiao}}, \bibinfo {author} {\bibfnamefont {F.}~\bibnamefont
  {Thöle}}, \bibinfo {author} {\bibfnamefont {S.~S.}\ \bibnamefont {Tsirkin}},
  \bibinfo {author} {\bibfnamefont {M.}~\bibnamefont {Wierzbowska}}, \bibinfo
  {author} {\bibfnamefont {N.}~\bibnamefont {Marzari}}, \bibinfo {author}
  {\bibfnamefont {D.}~\bibnamefont {Vanderbilt}}, \bibinfo {author}
  {\bibfnamefont {I.}~\bibnamefont {Souza}}, \bibinfo {author} {\bibfnamefont
  {A.~A.}\ \bibnamefont {Mostofi}},\ and\ \bibinfo {author} {\bibfnamefont
  {J.~R.}\ \bibnamefont {Yates}},\ }\bibfield  {title} {\bibinfo {title}
  {Wannier90 as a community code: new features and applications},\ }\href
  {https://doi.org/10.1088/1361-648x/ab51ff} {\bibfield  {journal} {\bibinfo
  {journal} {J. Condens. Matter Phys.}\ }\textbf {\bibinfo {volume} {32}},\
  \bibinfo {pages} {165902} (\bibinfo {year} {2020})}\BibitemShut {NoStop}%
\bibitem [{\citenamefont {Ozaki}(2003)}]{ozaki_openmx_2003}%
  \BibitemOpen
  \bibfield  {author} {\bibinfo {author} {\bibfnamefont {T.}~\bibnamefont
  {Ozaki}},\ }\bibfield  {title} {\bibinfo {title} {Variationally optimized
  atomic orbitals for large-scale electronic structures},\ }\href
  {https://doi.org/10.1103/PhysRevB.67.155108} {\bibfield  {journal} {\bibinfo
  {journal} {Phys. Rev. B}\ }\textbf {\bibinfo {volume} {67}},\ \bibinfo
  {pages} {155108} (\bibinfo {year} {2003})}\BibitemShut {NoStop}%
\bibitem [{\citenamefont {Ozaki}\ and\ \citenamefont
  {Kino}(2004)}]{ozaki_openmx_2004}%
  \BibitemOpen
  \bibfield  {author} {\bibinfo {author} {\bibfnamefont {T.}~\bibnamefont
  {Ozaki}}\ and\ \bibinfo {author} {\bibfnamefont {H.}~\bibnamefont {Kino}},\
  }\bibfield  {title} {\bibinfo {title} {Numerical atomic basis orbitals from
  {H} to {Kr}},\ }\href {https://doi.org/10.1103/PhysRevB.69.195113} {\bibfield
   {journal} {\bibinfo  {journal} {Phys. Rev. B}\ }\textbf {\bibinfo {volume}
  {69}},\ \bibinfo {pages} {195113} (\bibinfo {year} {2004})}\BibitemShut
  {NoStop}%
\bibitem [{\citenamefont {Isobe}\ and\ \citenamefont
  {Nagaosa}(2015)}]{isobe_enhancement_2015}%
  \BibitemOpen
  \bibfield  {author} {\bibinfo {author} {\bibfnamefont {H.}~\bibnamefont
  {Isobe}}\ and\ \bibinfo {author} {\bibfnamefont {N.}~\bibnamefont
  {Nagaosa}},\ }\bibfield  {title} {\bibinfo {title} {Enhancement of spin-orbit
  interaction by competition between {Hund}'s coupling and electron hopping},\
  }\href {https://doi.org/10.1088/1742-6596/592/1/012058} {\bibfield  {journal}
  {\bibinfo  {journal} {J. Phys.: Conf. Ser.}\ }\textbf {\bibinfo {volume}
  {592}},\ \bibinfo {pages} {012058} (\bibinfo {year} {2015})}\BibitemShut
  {NoStop}%
\bibitem [{\citenamefont {Tamai}\ \emph {et~al.}(2019)\citenamefont {Tamai},
  \citenamefont {Zingl}, \citenamefont {Rozbicki}, \citenamefont {Cappelli},
  \citenamefont {Riccò}, \citenamefont {de~la Torre}, \citenamefont
  {McKeown~Walker}, \citenamefont {Bruno}, \citenamefont {King}, \citenamefont
  {Meevasana}, \citenamefont {Shi}, \citenamefont {Radović}, \citenamefont
  {Plumb}, \citenamefont {Gibbs}, \citenamefont {Mackenzie}, \citenamefont
  {Berthod}, \citenamefont {Strand}, \citenamefont {Kim}, \citenamefont
  {Georges},\ and\ \citenamefont {Baumberger}}]{tamai_SRO_2019}%
  \BibitemOpen
  \bibfield  {author} {\bibinfo {author} {\bibfnamefont {A.}~\bibnamefont
  {Tamai}}, \bibinfo {author} {\bibfnamefont {M.}~\bibnamefont {Zingl}},
  \bibinfo {author} {\bibfnamefont {E.}~\bibnamefont {Rozbicki}}, \bibinfo
  {author} {\bibfnamefont {E.}~\bibnamefont {Cappelli}}, \bibinfo {author}
  {\bibfnamefont {S.}~\bibnamefont {Riccò}}, \bibinfo {author} {\bibfnamefont
  {A.}~\bibnamefont {de~la Torre}}, \bibinfo {author} {\bibfnamefont
  {S.}~\bibnamefont {McKeown~Walker}}, \bibinfo {author} {\bibfnamefont
  {F.}~\bibnamefont {Bruno}}, \bibinfo {author} {\bibfnamefont
  {P.}~\bibnamefont {King}}, \bibinfo {author} {\bibfnamefont {W.}~\bibnamefont
  {Meevasana}}, \bibinfo {author} {\bibfnamefont {M.}~\bibnamefont {Shi}},
  \bibinfo {author} {\bibfnamefont {M.}~\bibnamefont {Radović}}, \bibinfo
  {author} {\bibfnamefont {N.}~\bibnamefont {Plumb}}, \bibinfo {author}
  {\bibfnamefont {A.}~\bibnamefont {Gibbs}}, \bibinfo {author} {\bibfnamefont
  {A.}~\bibnamefont {Mackenzie}}, \bibinfo {author} {\bibfnamefont
  {C.}~\bibnamefont {Berthod}}, \bibinfo {author} {\bibfnamefont
  {H.}~\bibnamefont {Strand}}, \bibinfo {author} {\bibfnamefont
  {M.}~\bibnamefont {Kim}}, \bibinfo {author} {\bibfnamefont {A.}~\bibnamefont
  {Georges}},\ and\ \bibinfo {author} {\bibfnamefont {F.}~\bibnamefont
  {Baumberger}},\ }\bibfield  {title} {\bibinfo {title} {High-{Resolution}
  {Photoemission} on {Sr$_2$RuO$_4$} {Reveals} {Correlation}-{Enhanced}
  {Effective} {Spin}-{Orbit} {Coupling} and {Dominantly} {Local}
  {Self}-{Energies}},\ }\href {https://doi.org/10.1103/PhysRevX.9.021048}
  {\bibfield  {journal} {\bibinfo  {journal} {Phys. Rev. X}\ }\textbf {\bibinfo
  {volume} {9}},\ \bibinfo {pages} {021048} (\bibinfo {year}
  {2019})}\BibitemShut {NoStop}%
\bibitem [{\citenamefont {Cable}\ \emph {et~al.}(1961)\citenamefont {Cable},
  \citenamefont {Wilkinson},\ and\ \citenamefont {Wollan}}]{cable_CrCl3_1961}%
  \BibitemOpen
  \bibfield  {author} {\bibinfo {author} {\bibfnamefont {J.}~\bibnamefont
  {Cable}}, \bibinfo {author} {\bibfnamefont {M.}~\bibnamefont {Wilkinson}},\
  and\ \bibinfo {author} {\bibfnamefont {E.}~\bibnamefont {Wollan}},\
  }\bibfield  {title} {\bibinfo {title} {Neutron diffraction investigation of
  antiferromagnetism in $\mathrm{CrCl}_3$},\ }\href
  {https://doi.org/https://doi.org/10.1016/0022-3697(61)90053-1} {\bibfield
  {journal} {\bibinfo  {journal} {J. Phys. Chem. Solids}\ }\textbf {\bibinfo
  {volume} {19}},\ \bibinfo {pages} {29} (\bibinfo {year} {1961})}\BibitemShut
  {NoStop}%
\bibitem [{\citenamefont {McGuire}\ \emph {et~al.}(2017)\citenamefont
  {McGuire}, \citenamefont {Clark}, \citenamefont {KC}, \citenamefont {Chance},
  \citenamefont {Jellison}, \citenamefont {Cooper}, \citenamefont {Xu},\ and\
  \citenamefont {Sales}}]{mcguire_CrCl3_2017}%
  \BibitemOpen
  \bibfield  {author} {\bibinfo {author} {\bibfnamefont {M.~A.}\ \bibnamefont
  {McGuire}}, \bibinfo {author} {\bibfnamefont {G.}~\bibnamefont {Clark}},
  \bibinfo {author} {\bibfnamefont {S.}~\bibnamefont {KC}}, \bibinfo {author}
  {\bibfnamefont {W.~M.}\ \bibnamefont {Chance}}, \bibinfo {author}
  {\bibfnamefont {G.~E.}\ \bibnamefont {Jellison}}, \bibinfo {author}
  {\bibfnamefont {V.~R.}\ \bibnamefont {Cooper}}, \bibinfo {author}
  {\bibfnamefont {X.}~\bibnamefont {Xu}},\ and\ \bibinfo {author}
  {\bibfnamefont {B.~C.}\ \bibnamefont {Sales}},\ }\bibfield  {title} {\bibinfo
  {title} {Magnetic behavior and spin-lattice coupling in cleavable van der
  {Waals} layered {CrCl$_3$} crystals},\ }\href
  {https://doi.org/10.1103/PhysRevMaterials.1.014001} {\bibfield  {journal}
  {\bibinfo  {journal} {Phys. Rev. Mater.}\ }\textbf {\bibinfo {volume} {1}},\
  \bibinfo {pages} {014001} (\bibinfo {year} {2017})}\BibitemShut {NoStop}%
\bibitem [{\citenamefont {Dillon}\ and\ \citenamefont
  {Olson}(1965)}]{dillon_CrI3_1965}%
  \BibitemOpen
  \bibfield  {author} {\bibinfo {author} {\bibfnamefont {J.~F.}\ \bibnamefont
  {Dillon}}\ and\ \bibinfo {author} {\bibfnamefont {C.~E.}\ \bibnamefont
  {Olson}},\ }\bibfield  {title} {\bibinfo {title} {{Magnetization},
  {Resonance}, and {Optical} {Properties} of the {Ferromagnet} {CrI$_3$}},\
  }\href {https://doi.org/10.1063/1.1714194} {\bibfield  {journal} {\bibinfo
  {journal} {J. Appl. Phys.}\ }\textbf {\bibinfo {volume} {36}},\ \bibinfo
  {pages} {1259} (\bibinfo {year} {1965})}\BibitemShut {NoStop}%
\bibitem [{\citenamefont {Badrtdinov}\ \emph {et~al.}(2021)\citenamefont
  {Badrtdinov}, \citenamefont {Ding}, \citenamefont {Ritter}, \citenamefont
  {Hembacher}, \citenamefont {Ahmed}, \citenamefont {Skourski},\ and\
  \citenamefont {Tsirlin}}]{badrtdinov_mop_3sio_11_2021}%
  \BibitemOpen
  \bibfield  {author} {\bibinfo {author} {\bibfnamefont {D.~I.}\ \bibnamefont
  {Badrtdinov}}, \bibinfo {author} {\bibfnamefont {L.}~\bibnamefont {Ding}},
  \bibinfo {author} {\bibfnamefont {C.}~\bibnamefont {Ritter}}, \bibinfo
  {author} {\bibfnamefont {J.}~\bibnamefont {Hembacher}}, \bibinfo {author}
  {\bibfnamefont {N.}~\bibnamefont {Ahmed}}, \bibinfo {author} {\bibfnamefont
  {Y.}~\bibnamefont {Skourski}},\ and\ \bibinfo {author} {\bibfnamefont
  {A.~A.}\ \bibnamefont {Tsirlin}},\ }\bibfield  {title} {\bibinfo {title}
  {$\mathrm{MoP}_{3}\mathrm{SiO}_{11}$: A $4{d}^{3}$ honeycomb antiferromagnet
  with disconnected octahedra},\ }\href
  {https://doi.org/10.1103/PhysRevB.104.094428} {\bibfield  {journal} {\bibinfo
   {journal} {Phys. Rev. B}\ }\textbf {\bibinfo {volume} {104}},\ \bibinfo
  {pages} {094428} (\bibinfo {year} {2021})}\BibitemShut {NoStop}%
\bibitem [{\citenamefont {Maharaj}\ \emph {et~al.}(2018)\citenamefont
  {Maharaj}, \citenamefont {Sala}, \citenamefont {Marjerrison}, \citenamefont
  {Stone}, \citenamefont {Greedan},\ and\ \citenamefont
  {Gaulin}}]{maharaj_La2LiOsO6_2018}%
  \BibitemOpen
  \bibfield  {author} {\bibinfo {author} {\bibfnamefont {D.~D.}\ \bibnamefont
  {Maharaj}}, \bibinfo {author} {\bibfnamefont {G.}~\bibnamefont {Sala}},
  \bibinfo {author} {\bibfnamefont {C.~A.}\ \bibnamefont {Marjerrison}},
  \bibinfo {author} {\bibfnamefont {M.~B.}\ \bibnamefont {Stone}}, \bibinfo
  {author} {\bibfnamefont {J.~E.}\ \bibnamefont {Greedan}},\ and\ \bibinfo
  {author} {\bibfnamefont {B.~D.}\ \bibnamefont {Gaulin}},\ }\bibfield  {title}
  {\bibinfo {title} {Spin gaps in the ordered states of
  $\mathrm{La}_2\mathrm{LiXO}_6$($\mathrm{X}$=$\mathrm{Ru}$,$\mathrm{Os}$) and
  their relation to the distortion of the cubic double perovskite structure in
  $4d^3$ and $5d^3$ magnets},\ }\href
  {https://doi.org/10.1103/PhysRevB.98.104434} {\bibfield  {journal} {\bibinfo
  {journal} {Phys. Rev. B}\ }\textbf {\bibinfo {volume} {98}},\ \bibinfo
  {pages} {104434} (\bibinfo {year} {2018})}\BibitemShut {NoStop}%
\bibitem [{\citenamefont {Kermarrec}\ \emph {et~al.}(2015)\citenamefont
  {Kermarrec}, \citenamefont {Marjerrison}, \citenamefont {Thompson},
  \citenamefont {Maharaj}, \citenamefont {Levin}, \citenamefont {Kroeker},
  \citenamefont {Granroth}, \citenamefont {Flacau}, \citenamefont {Yamani},
  \citenamefont {Greedan},\ and\ \citenamefont
  {Gaulin}}]{Kermarrec_ba2yoso6_2015}%
  \BibitemOpen
  \bibfield  {author} {\bibinfo {author} {\bibfnamefont {E.}~\bibnamefont
  {Kermarrec}}, \bibinfo {author} {\bibfnamefont {C.~A.}\ \bibnamefont
  {Marjerrison}}, \bibinfo {author} {\bibfnamefont {C.~M.}\ \bibnamefont
  {Thompson}}, \bibinfo {author} {\bibfnamefont {D.~D.}\ \bibnamefont
  {Maharaj}}, \bibinfo {author} {\bibfnamefont {K.}~\bibnamefont {Levin}},
  \bibinfo {author} {\bibfnamefont {S.}~\bibnamefont {Kroeker}}, \bibinfo
  {author} {\bibfnamefont {G.~E.}\ \bibnamefont {Granroth}}, \bibinfo {author}
  {\bibfnamefont {R.}~\bibnamefont {Flacau}}, \bibinfo {author} {\bibfnamefont
  {Z.}~\bibnamefont {Yamani}}, \bibinfo {author} {\bibfnamefont {J.~E.}\
  \bibnamefont {Greedan}},\ and\ \bibinfo {author} {\bibfnamefont {B.~D.}\
  \bibnamefont {Gaulin}},\ }\bibfield  {title} {\bibinfo {title} {Frustrated
  fcc antiferromagnet {${\mathrm{Ba}}_{2}{\mathrm{YOsO}}_{6}$}: Structural
  characterization, magnetic properties, and neutron scattering studies},\
  }\href {https://doi.org/10.1103/PhysRevB.91.075133} {\bibfield  {journal}
  {\bibinfo  {journal} {Phys. Rev. B}\ }\textbf {\bibinfo {volume} {91}},\
  \bibinfo {pages} {075133} (\bibinfo {year} {2015})}\BibitemShut {NoStop}%
\end{thebibliography}%

\end{document}